\documentclass[twocolumn]{aastex63}

\usepackage{graphicx}
\usepackage{verbatim}
\usepackage{outlines}

\usepackage{amssymb}
\usepackage{color}
\usepackage{amsmath,amstext,amssymb}
\usepackage{enumitem}
\usepackage[mathscr]{eucal}
\usepackage[T1]{fontenc}
\usepackage{amsmath}

\newcommand{\needcite}[1]{{\color{blue}[??]}}
\newcommand{\amp}{A_{\rm GWB}}
\newcommand{\gwbamp}[1]{$\amp = #1 \times 10^{-15}$}

\newcommand{\sn}{$\hat{\rho}$}
\newcommand{\rhohd}{$\rho_\mathrm{HD}$}

\begin{document}

\title{Astrophysics Milestones For Pulsar Timing Array Gravitational Wave Detection}

\shorttitle{Astrophysics Milestones for PTAs}


\author[0000-0002-8826-1285]{Nihan S. Pol}
\affiliation{Department of Physics and Astronomy, Vanderbilt University, 2301 Vanderbilt Place, Nashville, TN 37235, USA}
\affiliation{Department of Physics and Astronomy, West Virginia University, P.O. Box 6315, Morgantown, WV 26506, USA}
\affiliation{Center for Gravitational Waves and Cosmology, West Virginia University, Chestnut Ridge Research Building, Morgantown, WV 26505, USA}
\author[0000-0003-0264-1453]{Stephen R. Taylor}
\affiliation{Department of Physics and Astronomy, Vanderbilt University, 2301 Vanderbilt Place, Nashville, TN 37235, USA}
\author[0000-0002-6625-6450]{Luke Zoltan Kelley}
\affiliation{Center for Interdisciplinary Exploration and Research in Astrophysics (CIERA), Northwestern University, Evanston, IL 60208}
\author[0000-0003-4700-9072]{Sarah J. Vigeland}
\affiliation{Center for Gravitation, Cosmology and Astrophysics, Department of Physics, University of Wisconsin-Milwaukee,\\ P.O. Box 413, Milwaukee, WI 53201, USA}
\author[0000-0003-1407-6607]{Joseph Simon}
\affiliation{Jet Propulsion Laboratory, California Institute of Technology, 4800 Oak Grove Drive, Pasadena, CA 91109, USA}
\affiliation{Department of Astrophysical and Planetary Sciences, University of Colorado, Boulder, CO 80309, USA}
\author[0000-0002-3118-5963]{Siyuan Chen}
\affiliation{Laboratoire de Physique et Chimie de l'Environment et de l'Espace, LPC2E UMR7328, Universite d'Orleans, CNRS, 45071 Orleans, France}
\affiliation{Station de Radioastronomie de Nancay, Observatoire de Paris, Universite PSL, CNRS, Universite d'Orleans, 18330 Nancay, France}
\affiliation{FEMTO-ST Institut de recherche, Department of Time and Frequency, UBFC and CNRS, ENSMM, 25030 Besancon, France}
\author{Zaven Arzoumanian}
\affiliation{X-Ray Astrophysics Laboratory, NASA Goddard Space Flight Center, Code 662, Greenbelt, MD 20771, USA}
\author[0000-0003-2745-753X]{Paul T. Baker}
\affiliation{Department of Physics and Astronomy, Widener University, One University Place, Chester, PA 19013, USA}
\author[0000-0003-0909-5563]{Bence B\'{e}csy}
\affiliation{Department of Physics, Montana State University, Bozeman, MT 59717, USA}
\author{Adam Brazier}
\affiliation{Cornell Center for Astrophysics and Planetary Science and Department of Astronomy, Cornell University, Ithaca, NY 14853, USA}
\author[0000-0003-3053-6538]{Paul R. Brook}
\affiliation{Department of Physics and Astronomy, West Virginia University, P.O. Box 6315, Morgantown, WV 26506, USA}
\affiliation{Center for Gravitational Waves and Cosmology, West Virginia University, Chestnut Ridge Research Building, Morgantown, WV 26505, USA}
\author[0000-0003-4052-7838]{Sarah Burke-Spolaor}
\affiliation{Department of Physics and Astronomy, West Virginia University, P.O. Box 6315, Morgantown, WV 26506, USA}
\affiliation{Center for Gravitational Waves and Cosmology, West Virginia University, Chestnut Ridge Research Building, Morgantown, WV 26505, USA}
\affiliation{CIFAR Azrieli Global Scholars program, CIFAR, Toronto, Canada}
\author[0000-0002-2878-1502]{Shami Chatterjee}
\affiliation{Cornell Center for Astrophysics and Planetary Science and Department of Astronomy, Cornell University, Ithaca, NY 14853, USA}
\author[0000-0002-4049-1882]{James M. Cordes}
\affiliation{Cornell Center for Astrophysics and Planetary Science and Department of Astronomy, Cornell University, Ithaca, NY 14853, USA}
\author[0000-0002-7435-0869]{Neil J. Cornish}
\affiliation{Department of Physics, Montana State University, Bozeman, MT 59717, USA}
\author[0000-0002-2578-0360]{Fronefield Crawford}
\affiliation{Department of Physics and Astronomy, Franklin \& Marshall College, P.O. Box 3003, Lancaster, PA 17604, USA}
\author[0000-0002-6039-692X]{H. Thankful Cromartie}
\affiliation{University of Virginia, Department of Astronomy, P.O. Box 400325, Charlottesville, VA 22904, USA}
\author[0000-0002-2185-1790]{Megan E. DeCesar}
\altaffiliation{NANOGrav Physics Frontiers Center Postdoctoral Fellow}
\affiliation{Department of Physics, Lafayette College, Easton, PA 18042, USA}
\affiliation{AAAS, STPF, ORISE Fellow hosted by the U.S. Department of Energy}
\author[0000-0002-6664-965X]{Paul B. Demorest}
\affiliation{National Radio Astronomy Observatory, 1003 Lopezville Rd., Socorro, NM 87801, USA}
\author[0000-0001-8885-6388]{Timothy Dolch}
\affiliation{Department of Physics, Hillsdale College, 33 E. College Street, Hillsdale, MI 49242, USA}
\affiliation{Eureka Scientific, Inc.  2452 Delmer Street, Suite 100, Oakland, CA 94602-3017}
\author{Elizabeth C. Ferrara}
\affiliation{NASA Goddard Space Flight Center, Greenbelt, MD 20771, USA}
\author[0000-0001-5645-5336]{William Fiore}
\affiliation{Department of Physics and Astronomy, West Virginia University, P.O. Box 6315, Morgantown, WV 26506, USA}
\affiliation{Center for Gravitational Waves and Cosmology, West Virginia University, Chestnut Ridge Research Building, Morgantown, WV 26505, USA}
\author[0000-0001-8384-5049]{Emmanuel Fonseca}
\affiliation{Department of Physics, McGill University, 3600  University St., Montreal, QC H3A 2T8, Canada}
\author[0000-0001-6166-9646]{Nathan Garver-Daniels}
\affiliation{Department of Physics and Astronomy, West Virginia University, P.O. Box 6315, Morgantown, WV 26506, USA}
\affiliation{Center for Gravitational Waves and Cosmology, West Virginia University, Chestnut Ridge Research Building, Morgantown, WV 26505, USA}
\author[0000-0003-1884-348X]{Deborah C. Good}
\affiliation{Department of Physics and Astronomy, University of British Columbia, 6224 Agricultural Road, Vancouver, BC V6T 1Z1, Canada}
\author[0000-0003-2742-3321]{Jeffrey S. Hazboun}
\altaffiliation{NANOGrav Physics Frontiers Center Postdoctoral Fellow}
\affiliation{University of Washington Bothell, 18115 Campus Way NE, Bothell, WA 98011, USA}
\author[0000-0003-1082-2342]{Ross J. Jennings}
\affiliation{Cornell Center for Astrophysics and Planetary Science and Department of Astronomy, Cornell University, Ithaca, NY 14853, USA}
\author[0000-0001-6607-3710]{Megan L. Jones}
\affiliation{Center for Gravitation, Cosmology and Astrophysics, Department of Physics, University of Wisconsin-Milwaukee,\\ P.O. Box 413, Milwaukee, WI 53201, USA}
\author[0000-0002-3654-980X]{Andrew R. Kaiser}
\affiliation{Department of Physics and Astronomy, West Virginia University, P.O. Box 6315, Morgantown, WV 26506, USA}
\affiliation{Center for Gravitational Waves and Cosmology, West Virginia University, Chestnut Ridge Research Building, Morgantown, WV 26505, USA}
\author[0000-0001-6295-2881]{David L. Kaplan}
\affiliation{Center for Gravitation, Cosmology and Astrophysics, Department of Physics, University of Wisconsin-Milwaukee,\\ P.O. Box 413, Milwaukee, WI 53201, USA}
\author[0000-0003-0123-7600]{Joey Shapiro Key}
\affiliation{University of Washington Bothell, 18115 Campus Way NE, Bothell, WA 98011, USA}
\author[0000-0003-0721-651X]{Michael T. Lam}
\affiliation{School of Physics and Astronomy, Rochester Institute of Technology, Rochester, NY 14623, USA}
\affiliation{Laboratory for Multiwavelength Astronomy, Rochester Institute of Technology, Rochester, NY 14623, USA}
\author{T. Joseph W. Lazio}
\affiliation{Jet Propulsion Laboratory, California Institute of Technology, 4800 Oak Grove Drive, Pasadena, CA 91109, USA}
\author{Jing Luo}
\affiliation{Department of Astronomy \& Astrophysics, University of Toronto, 50 Saint George Street, Toronto, ON M5S 3H4, Canada}
\author[0000-0001-5229-7430]{Ryan S. Lynch}
\affiliation{Green Bank Observatory, P.O. Box 2, Green Bank, WV 24944, USA}
\author[0000-0003-2285-0404]{Dustin R. Madison}
\altaffiliation{NANOGrav Physics Frontiers Center Postdoctoral Fellow}
\affiliation{Department of Physics and Astronomy, West Virginia University, P.O. Box 6315, Morgantown, WV 26506, USA}
\affiliation{Center for Gravitational Waves and Cosmology, West Virginia University, Chestnut Ridge Research Building, Morgantown, WV 26505, USA}
\author{Alexander McEwen}
\affiliation{Center for Gravitation, Cosmology and Astrophysics, Department of Physics, University of Wisconsin-Milwaukee,\\ P.O. Box 413, Milwaukee, WI 53201, USA}
\author[0000-0001-7697-7422]{Maura A. McLaughlin}
\affiliation{Department of Physics and Astronomy, West Virginia University, P.O. Box 6315, Morgantown, WV 26506, USA}
\affiliation{Center for Gravitational Waves and Cosmology, West Virginia University, Chestnut Ridge Research Building, Morgantown, WV 26505, USA}
\author[0000-0002-4307-1322]{Chiara M. F. Mingarelli}
\affiliation{Center for Computational Astrophysics, Flatiron Institute, 162 5th Avenue, New York, New York, 10010, USA}
\affiliation{Department of Physics, University of Connecticut, 196 Auditorium Road, U-3046, Storrs, CT 06269-3046, USA}
\author[0000-0002-3616-5160]{Cherry Ng}
\affiliation{Dunlap Institute for Astronomy and Astrophysics, University of Toronto, 50 St. George St., Toronto, ON M5S 3H4, Canada}
\author[0000-0002-6709-2566]{David J. Nice}
\affiliation{Department of Physics, Lafayette College, Easton, PA 18042, USA}
\author[0000-0001-5465-2889]{Timothy T. Pennucci}
\altaffiliation{NANOGrav Physics Frontiers Center Postdoctoral Fellow}
\affiliation{National Radio Astronomy Observatory, 520 Edgemont Road, Charlottesville, VA 22903, USA}
\affiliation{Institute of Physics, E\"{o}tv\"{o}s Lor\'{a}nd University, P\'{a}zm\'{a}ny P. s. 1/A, 1117 Budapest, Hungary}
\author[0000-0001-5799-9714]{Scott M. Ransom}
\affiliation{National Radio Astronomy Observatory, 520 Edgemont Road, Charlottesville, VA 22903, USA}
\author[0000-0002-5297-5278]{Paul S. Ray}
\affiliation{Space Science Division, Naval Research Laboratory, Washington, DC 20375-5352, USA}
\author[0000-0002-7283-1124]{Brent J. Shapiro-Albert}
\affiliation{Department of Physics and Astronomy, West Virginia University, P.O. Box 6315, Morgantown, WV 26506, USA}
\affiliation{Center for Gravitational Waves and Cosmology, West Virginia University, Chestnut Ridge Research Building, Morgantown, WV 26505, USA}
\author[0000-0002-7778-2990]{Xavier Siemens}
\affiliation{Department of Physics, Oregon State University, Corvallis, OR 97331, USA}
\affiliation{Center for Gravitation, Cosmology and Astrophysics, Department of Physics, University of Wisconsin-Milwaukee,\\ P.O. Box 413, Milwaukee, WI 53201, USA}
\author[0000-0001-9784-8670]{Ingrid H. Stairs}
\affiliation{Department of Physics and Astronomy, University of British Columbia, 6224 Agricultural Road, Vancouver, BC V6T 1Z1, Canada}
\author[0000-0002-1797-3277]{Daniel R. Stinebring}
\affiliation{Department of Physics and Astronomy, Oberlin College, Oberlin, OH 44074, USA}
\author[0000-0002-1075-3837]{Joseph K. Swiggum}
\altaffiliation{NANOGrav Physics Frontiers Center Postdoctoral Fellow}
\affiliation{Department of Physics, Lafayette College, Easton, PA 18042, USA}
\author[0000-0002-4162-0033]{Michele Vallisneri}
\affiliation{Jet Propulsion Laboratory, California Institute of Technology, 4800 Oak Grove Drive, Pasadena, CA 91109, USA}
\author[0000-0001-9678-0299]{Haley Wahl}
\affiliation{Department of Physics and Astronomy, West Virginia University, P.O. Box 6315, Morgantown, WV 26506, USA}
\affiliation{Center for Gravitational Waves and Cosmology, West Virginia University, Chestnut Ridge Research Building, Morgantown, WV 26505, USA}
\author[0000-0002-6020-9274]{Caitlin A. Witt}
\affiliation{Department of Physics and Astronomy, West Virginia University, P.O. Box 6315, Morgantown, WV 26506, USA}
\affiliation{Center for Gravitational Waves and Cosmology, West Virginia University, Chestnut Ridge Research Building, Morgantown, WV 26505, USA}
\shortauthors{Pol et al.}

\correspondingauthor{Nihan Pol}
\email{nihan.pol@nanograv.org}

\collaboration{1000}{The NANOGrav Collaboration}
\noaffiliation

\begin{abstract}
    
    The NANOGrav Collaboration reported strong Bayesian evidence for a common-spectrum stochastic process in its 12.5-yr pulsar timing array dataset, with median characteristic strain amplitude at periods of a year of $A_{\rm yr} = 1.92^{+0.75}_{-0.55} \times 10^{-15}$. However, evidence for the quadrupolar Hellings \& Downs interpulsar correlations, which are characteristic of gravitational wave signals, was not yet significant.  We emulate and extend the NANOGrav dataset, injecting a wide range of stochastic gravitational wave background (GWB) signals that encompass a variety of amplitudes and spectral shapes, and quantify three key milestones:  
    (I) Given the amplitude measured in the 12.5 yr analysis and assuming this signal is a GWB, we expect to accumulate robust evidence of an interpulsar-correlated GWB signal with 15--17 yrs of data, i.e., an additional 2--5 yrs from the 12.5 yr dataset; (II) At the initial detection, we expect a fractional uncertainty of $40\%$ on the power-law strain spectrum slope, which is sufficient to distinguish a GWB of supermassive black-hole binary origin from some models predicting more exotic origins;
    (III) Similarly, the measured GWB amplitude will have an uncertainty of $44\%$ upon initial detection, allowing us to arbitrate between some population models of supermassive black-hole binaries. 
    In addition, power-law models are distinguishable from those having low-frequency spectral turnovers once 20~yrs of data are reached.
    Even though our study is based on the NANOGrav data, we also derive relations that allow for a generalization to other pulsar-timing array datasets. Most notably, by combining the data of individual arrays into the International Pulsar Timing Array, all of these milestones can be reached significantly earlier.
    
\end{abstract}


\section{Introduction}

Supermassive black hole binaries (SMBHB) are expected to form following the mergers of massive galaxies during the buildup of hierarchical structure in $\Lambda$CDM \citep[][and references therein]{2019A&ARv..27....5B}. No observationally-confirmed SMBHBs at sub-parsec separations are known, but a growing number of strong candidates have been identified \citep{2015MNRAS.453.1562G,2016MNRAS.463.2145C,2019ApJ...884...36L,2020arXiv200812329C}. If SMBHBs are indeed able to form, and also to reach $\sim$milli-parsec separations, then they will become the sources of the strongest gravitational waves (GWs) in the Universe.
The stochastic superposition of GWs produced by SMBHBs across cosmic time is expected to produce an aggregate background signal with an approximate characteristic-strain spectrum of the form $h_c(f) = A_\mathrm{GWB}(f/f_\mathrm{yr})^\alpha$ with $\alpha = -2/3$ for a continuous population of circular and purely GW-driven binary systems \citep{1995ApJ...446..543R,2003ApJ...583..616J,2008MNRAS.390..192S}. At low frequencies ($f \lesssim 10^{-8}$ Hz), interactions between SMBHBs and their ambient galactic environments can turn over or flatten the spectrum ($\alpha > -2/3$) \citep{2015PhRvD..91h4055S,2017PhRvL.118r1102T,2017MNRAS.471.4508K}. At high frequencies ($f \gtrsim 10^{-8}$ Hz)
rapid binary hardening can lead to a low occupation of spectral bins that steepens the spectrum ($\alpha < -2/3$) and may introduce spikes from individual sources \citep{2008MNRAS.390..192S}. Despite these effects, the stochastic gravitational wave background (GWB) from SMBHBs is expected to be the dominant source of GWs in the nanohertz frequency regime.

Pulsar Timing Arrays \citep{PTA_sazhin, PTA_detweiler, PTA_foster_backer} (PTAs) measure the times-of-arrival (TOAs) of radio pulses from millisecond pulsars as a means of measuring the local space-time curvature, and thus signs of passing GWs, analogous to how the LIGO-Virgo Collaboration \citep{LIGO_detector_ref, VIRGO_detector_ref} uses pairs of perpendicular laser arms. As PTAs accrue larger datasets over time, the increasing number of recorded pulses leads to improved sensitivity, especially as new pulsars are added to the arrays \citep{os_scaling_laws_siemens,2016ApJ...819L...6T}.
The NANOGrav $12.5$ yr dataset \citep{NG12p5_timing} shows conclusive evidence (with Bayesian odds of $\sim 10^4:1$) that a common-spectrum low-frequency stochastic process is present across a large fraction of the 45 millisecond pulsars included in the analysis \citep{NG12p5_gwb}. Its properties are consistent with expectations of the strain-spectrum from a stochastic GWB, but it does not yet show statistically-significant quadrupolar interpulsar correlations, which are widely considered the definitive evidence for the GWB \citep{HD_curve, tiburzi_spatial_corr}. The latter are the so-called `Hellings \& Downs' (HD) correlations of GWB-induced timing delays between pairs of pulsars, which are a function only of the angular separation between the pulsars on the sky.
Future datasets, such as the in-preparation NANOGrav $15$ yr dataset, in addition to data from the European PTA \citep[EPTA;][]{EPTA_dr1}, Australian Parkes PTA \citep[PPTA;][]{PPTA_dr2} and the International Pulsar Timing Array \citep[IPTA;][]{ipta_dr1, ipta_dr2}, will reveal the nature of the signal.
These data will either show spatial correlations consistent with a GWB, or other correlation signatures that suggest unmodeled noise processes affecting the PTA data.

In this paper, we forecast the evolution of GWB inference into the future, under the assumption that the signal is produced by cosmological SMBHBs. 
There may be alternate sources of the GWB, and there have been many recent suggestions of such interpretations for the NANOGrav 12.5~yr results \citep[see, e.g.,][]{ng12p5_pbh_1,ng12p5_pbh_2, ng12p5_cs_1,ng12p5_cs_2}. The scaling results that we derive here are generalizable to such other sources of the GWB. Similar analyses performed in the past were focused only on the time to detection of the GWB \citep[e.g.,][]{2016ApJ...819L...6T, rosado_forecast, sarah_forecast, Kelley+2017b}, and while this remains important (especially given the results in \citet{NG12p5_gwb}) we consider here for first time the evolution of the GWB parameter measurement uncertainties. 
The evolution of these properties is of crucial importance in determining the source and underlying physics that produces the GWB. For example, while a GWB from SMBHBs produces a spectral index of $\alpha = -2/3$, a GWB produced by primordial GWs can have a spectral index, $\alpha = -2 \textrm{ or } -1$ \citep[][]{primordial_gw_spectrum, lasky_pgw}, while a GWB produced by some models of cosmic strings have $\alpha = -7/6$ \citep[][]{cosmic_string_spectrum}. A distinction between these source models can only be made once the spectral index measurement uncertainty is small enough to exclude one (or more) of the predicted spectral indices. Similarly, different models of a given source can produce different amplitudes of the GWB, and thus knowing when we can distinguish between them is dependent on the evolution of the measurement uncertainties on the GWB amplitude.

However, for a background produced by inspiraling SMBHBs, the canonical $\alpha = -2/3$ spectral index GW strain power-law emerges when assuming a continuous distribution of circular binary sources evolving purely due to GW emission \citep{Phinney2001}.  Following galaxy mergers, two SMBHs can only become bound and reach the small separations ($\lesssim 10^{-2}$ pc) required to produce detectable GWs through environmental interactions, such as dynamical friction and stellar ``slingshot'' scattering \citep{Begelman+1980}.  These interactions make binaries harden faster than they would due purely to GW emission, and thus the resulting GW strain can be lower than in the $-2/3$ power-law, particular at low frequencies ($f \ll 1 \, \textrm{yr}^{-1}$).  At higher frequencies, finite number effects  \citep{2008MNRAS.390..192S} can also cause significant deviations from a pure power-law. We explore the effects of these variations in the SMBHB GWB spectrum by injecting and recovering ``realistic'' spectra that have been produced through full mock populations of SMBH binaries.

The paper is structured as follows. In Sec.~\ref{sec:methods} we describe the methods used to simulate and analyze the PTA datasets. In Sec.~\ref{sec:laplace_apprx}, we show that the Bayesian model odds ratio can be related to the frequentist signal-to-noise ratio through a simple analytic expression for PTAs, while in Sec.~\ref{sec:suff_stat} we derive a new statistic, the total signal-to-noise ratio, that encapsulates the information in both the auto- and cross-correlation components of the PTA data. In Sec.~\ref{sec:milestones} we describe the evolution of the GWB detection significance, parameter uncertainties, and spectral characterization of the GWB, and define three key milestones that PTAs should achieve along the way. Finally, in Sec.~\ref{sec:ipta}, we show how the timeline to achieving these milestones described can be accelerated through combining the data from individual PTAs into the IPTA.

\section{Methods} \label{sec:methods}

    \subsection{Simulation framework}
        
        We have developed a framework for realistic pulsar-timing data simulation that employs the NANOGrav 12.5 yr dataset \citep[][]{NG12p5_timing} as its foundation, deriving observational timestamps and TOA uncertainties from all $45$ pulsars therein, as well as useful meta-data such as observing radio frequency and telescope.
        We account for pulse-phase jitter noise, as well as the usual TOA uncertainties from radiometer noise, using the maximum likelihood pulsar noise parameters measured by NANOGrav in the 12.5 yr data preparation and analysis.
        We epoch-average the TOA uncertainties and timestamps which reduces the effective number of TOAs by almost an order of magnitude.
        We also inject intrinsic red noise in each pulsar according to its measured values in the NANOGrav 12.5 yr dataset, with the caveat that we filter out a GWB-like $\alpha = -2/3$ red process from each pulsar by modeling it alongside the instrinsic pulsar noise.
        This is important in isolating the true intrinsic red noise in each pulsar, which may otherwise be conflated with a common process in single-pulsar noise estimation. 
        These simulations are the first to inject red noise in each pulsar that reflects the measured NANOGrav common red noise being appropriately filtered out.
        
        In order to forecast data accumulation beyond the existing 12.5 yr baseline, we derive distributions of the observational cadences and TOA uncertainties from the final year's worth of data in each pulsar. We then adopt a status-quo approach of drawing new observation times and uncertainties from these distributions until we reach a $20$ year array baseline.
        We retain the 45 pulsars from the 12.5 yr analysis and do not add any new pulsars as a function of time.
        This latter choice implies that the temporal growth rate of the detection statistics that we derive in this work will be relatively conservative, since an addition of pulsars to the PTA is known to increase the detection significance of the GWB \citep{os_scaling_laws_siemens}.
        
        Finally, we inject ten different realizations of noise and isotropic GWB signals (with spectral index $\alpha = -2/3$) for each of eight different amplitudes in the range $10^{-16} \leq A_{\rm GWB} \leq 5 \times 10^{-15}$ into our synthesized $20$-year baseline array. We then study how the signal evolves in our data by analyzing increasingly longer slices of each simulation, from $10$ years up to the final $20$ years. Put together, our PTA simulation technique is the most sophisticated that has ever been presented in the literature.

    \subsection{Bayesian Data Analysis Methods}
        
        The methods used for modeling the injected GWB are the same as those used in the NANOGrav 11 yr \citep{NG11_sgwb} and 12.5 yr \citep{NG12p5_gwb} analyses. 
        Thus, we only briefly summarize the probabilistic framework for estimating the GWB parameters.
        
        In the standard detection pipeline, the GWB is modeled as a power-law across frequency for the characteristic strain,
        \begin{equation}
            \displaystyle h_{\rm c} (f) = A_{\rm GWB} \left( \frac{f}{1 \, {\rm yr}^{-1}} \right)^{\alpha},
            \label{eq:sgwb_powerlaw}
        \end{equation}
        where $A_{\rm GWB}$ is the amplitude of the GWB, $f$ is the frequency, and $\alpha = -2/3$ is the spectral index expected for a population of inspiraling SMBHBs whose inspiral is dominated by GWs. This GWB spectrum can be expressed in terms of the timing-residual cross-spectral density between pulsars $a$ and $b$ as,
        \begin{equation}
            \displaystyle S_{\rm ab} (f) = \Gamma_{\rm ab} \frac{A_{\rm GWB}^2}{12 \pi^2} \left( \frac{f}{1 \,{\rm yr}^{-1}} \right)^{-\gamma} {\rm yr^3},
            \label{eq:residual_sgwb_powerlaw}
        \end{equation}
        where $\gamma \equiv 3 - 2 \alpha$, and $\Gamma_{\rm ab}$ is the overlap reduction function that encodes the interpulsar spatial correlations, such as the HD spatial correlations \citep{HD_curve} expected for an isotropic GWB. 
        
        In addition to the GWB, we also need to model the intrinsic noise for each pulsar in the PTA. For each pulsar, we model uncorrelated instrumental Gaussian noise in the multiplicative EFAC and quadrature-additive EQUAD parameters. Noise that is correlated across observing frequencies but uncorrelated across TOAs is modeled using the ECORR parameter. These parameters are collectively referred to as ``white noise''. We include an additional power-law process per pulsar to model any intrinsic low-frequency noise (i.e. ``spin red noise'') in the given pulsar. The aforementioned white noise parameters that correct the reported TOA uncertainties were folded into the simulated uncertainties in the preparation of our datasets, so we do not separately model these white noise parameters in our analysis. The red noise for the pulsars is allowed to vary when we search for the GWB.
        
        The GWB parameter constraints are derived by modeling the GWB as a time-correlated process that is statistically common (i.e. with a common spectrum) but uncorrelated among all pulsars in the PTA. This ignores the influence of interpulsar correlations on GWB parameter constraints, which should be minimal as these are dominated by autocorrelation information in the pulsars. Constraints on the GWB spectral index, $\alpha$, are reported as marginalized over all parameters, including $A_\mathrm{GWB}$, while constraints on $A_\mathrm{GWB}$, are reported as marginalized over all noise parameters but conditioned on the fiducial spectral index $\alpha=-2/3$. The inclusion of HD correlations in the search process is computationally expensive, so we only use it to discriminate whether the observed common red process is from a GWB or other spatially correlated signals like terrestrial clock errors (monopolar spatial correlations) or Solar System ephemeris errors (dipolar spatial correlations). These spatial correlations are encoded in the overlap reduction function in Eq.~\ref{eq:residual_sgwb_powerlaw}. 
        
        To calculate the Bayesian detection significance, we use a product-space sampling method that computes the probabilistic preference for one signal model over another \citep[][and references therein]{NG11_sgwb,2020PhRvD.102h4039T}. We construct a super-model consisting of the union of multiple GWB signal models, where an indexing variable determines which model is ``active'' and used to calculate the corresponding likelihood. The ratio of the posterior probabilities of the model indices can then be used to estimate the Bayesian odds for the preference of one model over the other. Thus, to calculate the Bayesian odds for a HD-correlated process in the dataset, $\mathcal{O}_{\rm HD}$, we create a super-model with the same GWB parameters in both models, while the index parameter toggles the presence of HD correlations. 
        
        In addition to power-law strain spectrum models, we also measure the GWB strain spectrum agnostically with an independent amplitude at each frequency. Our priors correspond to independent log-uniform constraints on the characteristic strain amplitude for each Fourier-basis frequency, given by $k/T$, where $k = 1, 2, ..., 30$ and $T$ is the time-span between the first and last TOA in the given dataset. 
        Unless explicitly specified, all of the results presented in this paper have been derived using 30 frequencies to model the GWB (and intrinsic red noise) and employing the recent JPL solar system ephemeris, DE438 \citep[][]{de438}. However, we also explore the effect of using a different number of frequencies and BayesEphem \citep[][]{bayesephem} in modeling the GWB.
        
    \subsection{Frequentist Data Analysis Methods}
        
        In addition to the Bayesian pipeline, we also use a frequentist approach to model the GWB \citep{NG9_sgwb, NG11_sgwb}. This uses an optimal statistic, $\hat{A}^2$, to measure the amplitude of the GWB, corresponding to the sum of the correlations between pulsar pairs weighted by the intrinsic pulsar noise, and conditioned on the assumed spectral index of the GWB.
        The optimal statistic can also be used to calculate the signal-to-noise ratio, $\rho_{\rm template}$, for a given interpulsar correlation template, e.g., monopolar or dipolar for clock and ephemeris errors respectively, and HD (or quadrupolar) for an astrophysical GWB. To calculate the template S/N, the optimal statistic first calculates the amount of cross-correlated power between different pulsar pairs in the PTA. This pairwise cross-correlated power, suitably binned in angular separation space, can be used as a visual test for the presence of quadrupolar correlations in the data, or as intermediate products for further study of the spatial correlations. 
        
        In this work, we use the noise-marginalized optimal statistic \citep{NM_OS}, which is more robust when pulsars have intrinsic red noise. We use posterior samples from Bayesian MCMC analyses to marginalize over the pulsar intrinsic red noise and calculate corresponding distributions for $\hat{A}^2$ and $\rho_{\rm template}$.
        
        These Bayesian and frequentist methods reflect production-level analyses that are employed in PTA detection and parameter estimation analyses. Thus, our results are a crucial validation of the efficacy of the pipelines that will be performing the important business of SGWB detection in the next several years.

\section{Bridging Bayesian odds and frequentist signal-to-noise ratios} \label{sec:laplace_apprx}
    
    Under certain circumstances it is possible to relate Bayesian model selection to frequentist hypothesis testing. When data is informative such that the likelihood is strongly peaked, the Bayesian evidence can be computed under the \textit{Laplace approximation}, such that $Z_\mathcal{H}\equiv\int\,d\theta\, p(d|\theta,\mathcal{H})p(\theta|\mathcal{H})\approx p(\theta_\mathrm{ML}|\mathcal{H})\Delta V_\mathcal{H}/V_\mathcal{H}$, where $p(d|\theta_\mathrm{ML},\mathcal{H})$ maximizes the likelihood with parameters $\theta$ given data $d$ under model $\mathcal{H}$ \citep{2017LRR....20....2R}. $\Delta V_\mathcal{H}/V_\mathcal{H}$ measures the compactness of the parameter space volume occupied by the likelihood with respect to the total prior volume, incorporating the Bayesian notion of model parsimony. Taking the ratio of Bayesian evidences between two models, labeled 1 and 2, and assuming equal prior odds, allows the Bayesian odds ratio, $\mathcal{O}_{12}$, to be written as $\ln\mathcal{O}_{12}\approx\ln\Lambda_\mathrm{ML}(d) + \ln\left[ (\Delta V_1/V_1)/(\Delta V_2/V_2) \right]$, where $\Lambda_\mathrm{ML}(d)$ is the maximum likelihood ratio. The relevant maximum likelihood statistic for PTA GWB detection is the aforementioned \textit{optimal statistic}: a noise-weighted two-point correlation statistic between all unique pulsar pairs, which compares a model with HD interpulsar correlations versus one with no correlations between pulsars  \citep{abc+2009,dfg+13,ccs+2015}. The signal-to-noise ratio (S/N, $\rho$) of such GWB-induced correlations can be written as $\ln\Lambda_\mathrm{ML}=\rho^2/2$. While the likelihood may be marginally more compact under the model with HD correlations, there is no difference in parameter dimensionality; hence we ignore the likelihood compactness terms. The relationship between the Bayesian odds ratio in favor of HD correlations and the frequentist S/N of such correlations can then be written as,
    \begin{equation} \label{eq:laplace}
        \ln\mathcal{O}_\mathrm{HD} \approx \rho_{\rm HD}^2 / 2.
    \end{equation}
    
    \begin{figure}[htb]
    \centering
    \includegraphics[width = \columnwidth]{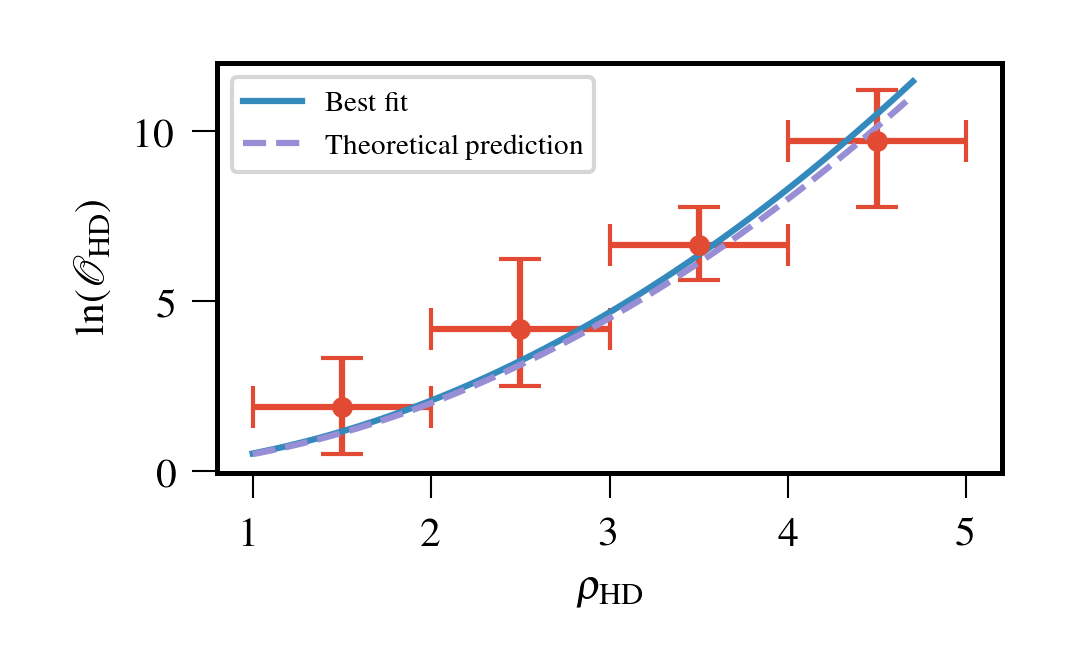}
    \caption{The Bayesian odds ratio preferring a GWB over a spatially uncorrelated process ($\mathcal{O}_\mathrm{HD}$) is plotted against the cross-correlation signal-to-noise ratio (S/N, $\rho$). These quantities were calculated using the injected simulated datasets across all realizations and injected amplitudes. The theoretical prediction (dashed blue line) for the relation between these two measures of confidence agrees with the empirical best-fit model (solid blue line).}
    \label{fig:30f_bf_v_snr}
\end{figure}
     
    In a bid to assess the validity of this relationship for PTA GWB searches, we used our simulated datasets to calculate both of these statistics across all realizations, injected amplitudes and time slices. These values, and the theoretical relationship in Eq.~\ref{eq:laplace}, are shown in Fig.~\ref{fig:30f_bf_v_snr}. We also allowed the pre-factor of $1/2$ to vary in an empirical fit, finding a value of $0.52 \pm 0.01$, which agrees well with Eq.~\ref{eq:laplace}. 
    This is the first time that this relationship has been validated for PTA GWB detection.
    
\section{A statistic to quantify PTA milestones} \label{sec:suff_stat}
    
    There are many factors that influence the detectability of a GWB signal in a given PTA configuration. Some of these are related to the signal itself, i.e.~the amplitude of the characteristic strain spectrum at $f=1/\mathrm{yr}$, $A_\mathrm{GWB}$, as well as the frequency-dependent shape of the spectrum. The quality of our detector is parametrized through the overall observational timing baseline of the array, the number of pulsars, and the timing quality of each pulsar (given by its radio-frequency dependent noise characteristics). We need a statistic that accounts for all of these factors, acting as a fiducial scaling parameter in terms of which we can track the evolution of parameter uncertainties and signal detectability. The optimal statistic is not appropriate here, since it only considers cross-correlations and thus underestimates the information content in the full signal (especially the auto-correlations) that dominate parameter estimation. 
    
    The full signal likelihood (modeling auto- and cross-correlations) is a sufficient statistic for this purpose. Hence we form a total signal S/N, $\hat\rho$, through the log-likelihood ratio of the GWB$+$noise versus noise-only models, where noise is treated as uncorrelated between pulsars. Therefore,
    \begin{equation}
        \hat\rho = \sqrt{2\ln\Lambda}\equiv \sqrt{2\left[\ln p(d|\vec\theta_\mathrm{signal},\vec\theta_\mathrm{noise}) - \ln p(d|\vec\theta_\mathrm{noise})\right]}.
        \label{eq:rho_hat_defn}
    \end{equation}
    This differs from HD cross-correlation S/N, \rhohd, (or an equivalent cross-correlation statistic) in considering all distinct pairings of the pulsars, including autocorrelations.
    
    We would like to note that this new statistic is not meant to be a replacement for the cross-correlation S/N. Since an astrophysical GWB is characterized by the unique HD cross-correlation signature, the corresponding cross-correlation S/N, \rhohd, will be the primary arbiter for confirming a signal as an astrophysical GWB.
    
    \begin{figure}[htb]
        \centering
        \includegraphics[width = \columnwidth]{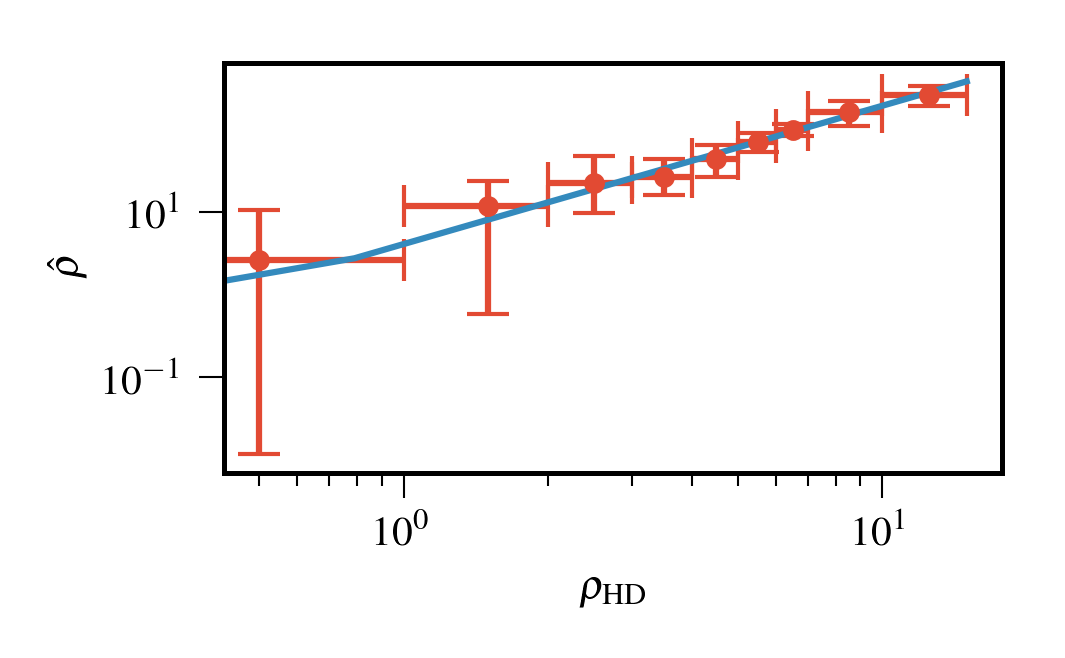}
        \caption{The evolution of the total S/N, \sn, as a function of the HD cross-correlation S/N, $\rho_{\rm HD}$. As we can see, HD template S/N values of 1, 3, and 7 correspond to \sn\ values of 5, 25 and 105 respectively.}
        \label{fig:rho_hat_v_hd_snr}
    \end{figure}

    As PTAs move closer to detection, the power in the GWB at low frequencies is expected to dominate the intrinsic white and red noise in some of the pulsars in the PTA. In this ``intermediate signal regime'', the total S/N, \sn, is expected to be much larger than the corresponding HD S/N \citep[\rhohd,][]{romano_acor_evn}. We can directly connect them by calculating the two statistics on the same simulated injected datasets. The binned values for the two statistics calculated this way are shown in Fig.~\ref{fig:rho_hat_v_hd_snr}, and related by the empirical scaling relation,
    \begin{equation}
        \displaystyle \hat{\rho} = 25 \, \left( \frac{\rho_{\scriptscriptstyle \rm HD}}{3} \right)^{1.7}.
        \label{eq:rho_hat_v_rho_hd}
    \end{equation}
    Using this scaling relation, \rhohd\ values of $[1, 3, 7]$ map to \sn\ values of approximately $[5, 25, 105]$ respectively. Since the power in the auto-correlations is significantly larger than in cross-correlations, the corresponding S/N values for the total SN, \sn, are higher than those for HD cross-correlation, $\rho_{\rm HD}$ \citep[][]{romano_acor_evn}.
    
    From existing scaling laws for $\rho_\mathrm{HD}$ \citep{os_scaling_laws_siemens}, we can generalize $\hat\rho$ to any PTA configuration. For a PTA in the weak and intermediate signal regime \citep{os_scaling_laws_siemens}, the total S/N can be written as,
    \begin{subequations}
    \begin{align}
        \! & \hat{\rho} \propto \left[ M c \frac{A_\mathrm{GWB}^2}{\sigma^2} T^{\gamma} \right]^{1.7} \label{eq:rho_hat_weak_regime} \\
         \! & \hat{\rho} \propto \left[ M \left( \frac{A_\mathrm{GWB}}{\sigma \sqrt{c}} \right)^{1/\gamma} T^{1/2} \right]^{1.7}, \label{eq:rho_hat_int_regime}
    \end{align}
    \label{eq:general_rho_hat}
    \end{subequations}
    respectively, where $M$ is the number of pulsars in the array, $c = 1 / \Delta t$ is the inverse of the observational cadence, $A_\mathrm{GWB}$ is the amplitude of the GWB, $\sigma$ is the white noise timing RMS, $T$ is the timing baseline of the PTA, and $\gamma$ is as previously defined ($\gamma\equiv 3-2\alpha=13/3$ for a population of SMBHBs).
    
\section{PTA Milestones} \label{sec:milestones}
    
\subsection{Milestone I: Detection}  

\begin{figure*}[htb]
    \centering
    \includegraphics[width=1.0\textwidth]{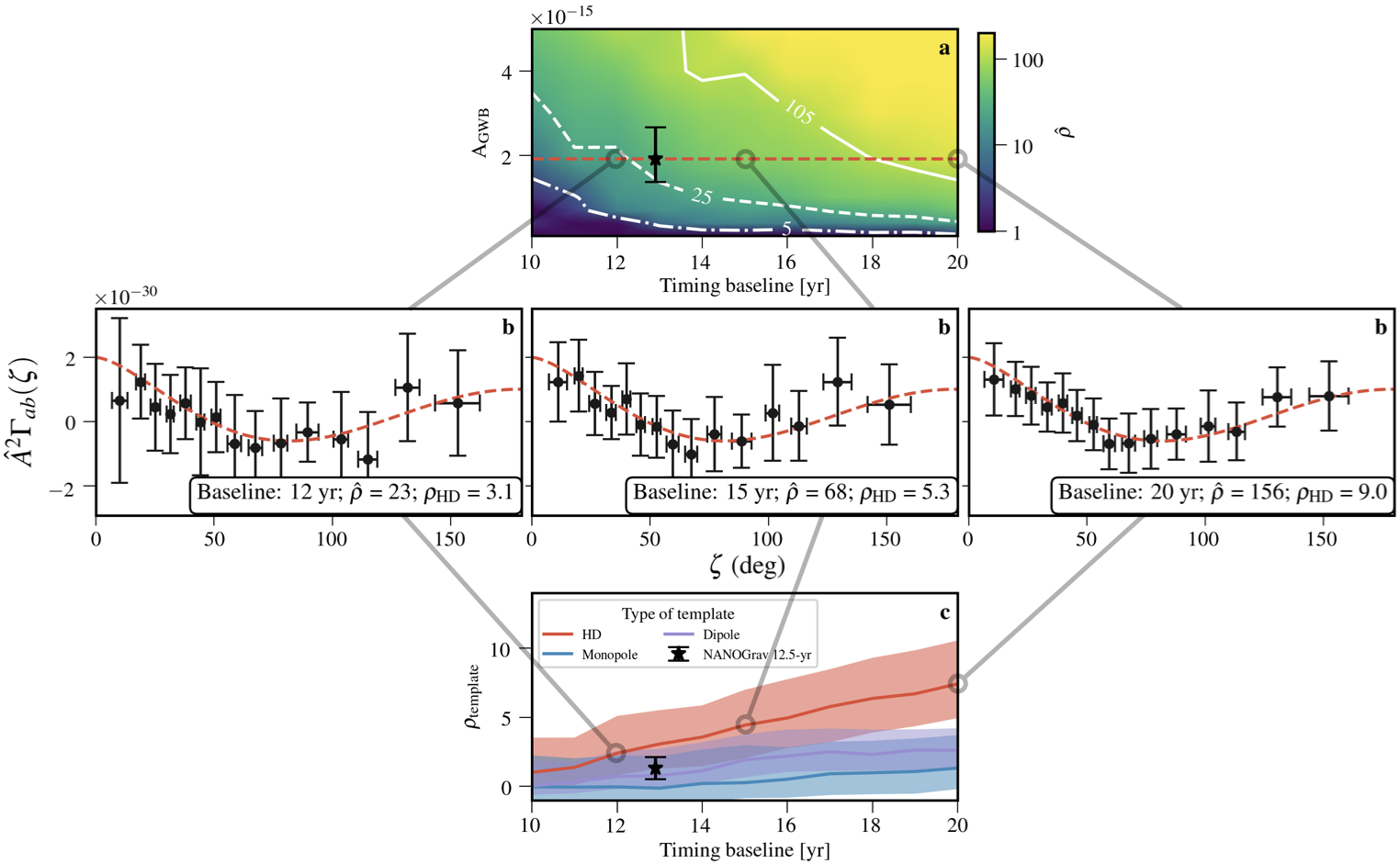}
    \caption{\textit{Top:} The evolution of the median total S/N, \sn, over ten realizations as a function of time. The dash-dot, dashed and solid contours represent total S/N values of 5, 25, and 105 corresponding to HD S/N, $\rho_{\rm HD}$, of 1, 3 and 7 respectively. The black data point represents the amplitude of the GWB from the NANOGrav 12.5 yr analysis.
    \textit{Middle:} The median binned cross-correlated power over ten realizations for increasing S/N from left to right. The uncertainties in this panel represent the spread of the cross-correlated power over the ten realizations. The theoretical HD curve for a power-law GWB with an amplitude of \gwbamp{2} is shown by the red dashed line.
    \textit{Bottom:} The cross-correlation S/N ratio, $\rho_{\rm template}$, for different types of spatial correlations with an injected power-law GWB amplitude of \gwbamp{[1-2]}.  The shaded regions represent the median 95\% credible intervals across ten realizations of the PTA dataset. The black data point represents the significance of the GWB from the NANOGrav 12.5 yr analysis.
    }
    \label{fig:forecast_snr}
\end{figure*}

The upper panel of Fig.~\ref{fig:forecast_snr} shows the evolution of the median $\hat\rho$ as a function of data baseline and injected GWB amplitude, the middle panels show the cross-correlated power in different angular separation bins for a GWB with an amplitude of \gwbamp{2}, for three different data baselines (12 yr, 15 yr, and 20 yr), and the lower panel shows the evolution of the S/N for HD, monopole, and dipole cross-correlations as a function of data baseline for a GWB with amplitude between \gwbamp{[1-2]}.
With an increase in the length of the PTA dataset, $\hat\rho$ for the GWB increases, as well as the S/N for HD correlations, $\rho_\mathrm{HD}$. The bottom panel of Fig.~\ref{fig:forecast_snr} shows that, starting at the 15 yr slice, the median HD correlations are preferred over other types of spatial correlation for a GWB with strength \gwbamp{[1-2]}. Only with HD correlations clearly favored over monopolar or dipolar signals (which happens near the 18 yr baseline in our simulations) can a signal be confidently attributed to GWBs instead of uncharacterized noise sources.

To test the influence of the number of frequencies used to model the GWB, we repeat the analysis in the lower panel of Fig.~\ref{fig:forecast_snr} for a GWB modeled with only the lowest $5$ frequencies.
The choice of these lowest 5 frequencies is designed to avoid GWB significance and parameter estimation being contaminated by poorly-modeled excess noise in the dataset that dominates at higher frequencies \citep{NG12p5_gwb}.
We find that the initial detection of a GWB is not inhibited or delayed by using only the lowest 5 frequency bins, which only reduces the S/N for HD correlations by a few percent. 
We also explored the influence that accounting for systematics in Solar-System ephemeris modeling has on our GWB detection significance. We do so by including \textsc{BayesEphem} \citep[][]{bayesephem} in our model, which marginalizes over uncertainties in Jupiter orbital elements, gas giant masses, and the celestial reference frame. We find that \textsc{BayesEphem} reduces detection significance by only $\sim4$ units of total S/N, \sn, consistent with earlier results reported in \citet{bayesephem}. Thus, the presence of solar system emphemeris uncertainties is unlikely to inhibit the detection of a GWB by PTAs.

As shown in Fig.~\ref{fig:forecast_snr} as a black data point, detection statistics for the NANOGrav 12.5 yr dataset are consistent with our simulations, and we expect the significance of a GWB to increase in the real dataset at least at the rate shown here, if not faster due to the addition of more pulsars to the array \citep{os_scaling_laws_siemens} and new advanced noise mitigate schemes that should allow the GWB spectrum to be modeled to higher frequencies.  

\begin{figure*}[htb]
    \centering
    \includegraphics[width = \textwidth]{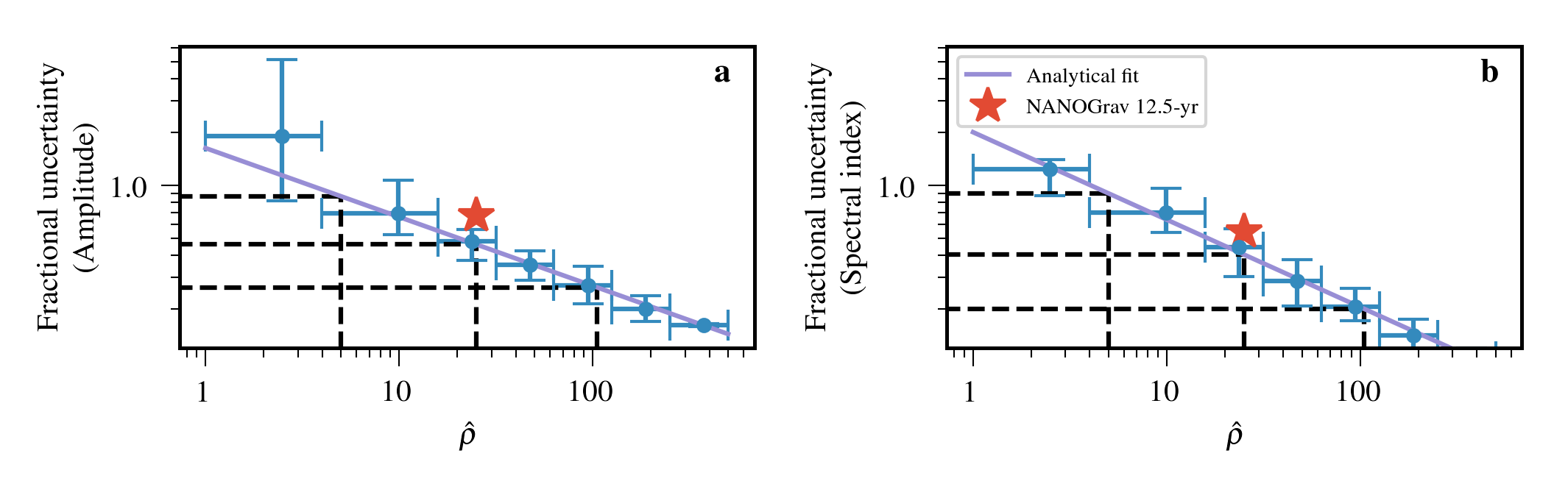}
    \caption{The evolution of the median fractional uncertainty on the measured amplitude, $A$, and spectral index, $\alpha$, as a function of total S/N, \sn{}, is shown in panels (a) and (b) respectively. The evolution of the fractional uncertainty for the amplitude and spectral index with the S/N is best fit by a power-law as shown (parametrized in Eq.~\ref{eq:func_scaling_law}). 
    For the amplitude and spectral index, the fractional uncertainties corresponding to \sn{} values of 5, 25 and 105 are 84\%, 44\% and 25\%, and 90\%, 40\% and 20\% respectively, and are shown by the dashed black lines. The star shows the fractional uncertainty from the NANOGrav 12.5 yr analysis. }
    \label{fig:forecast}
\end{figure*}

\subsection{Milestone II: Source Of The GWB} 

Fig.~\ref{fig:forecast} shows the evolution of the fractional parameter measurement uncertainty, $\Delta X/X$, as a function of total S/N, where $X$ is the median measured value of the GWB amplitude ($A_\mathrm{GWB}$) and spectral index ($\alpha$), and $\Delta X$ is the corresponding 95\% credible interval parameter uncertainty. These uncertainties are well fit by the relations,
\begin{subequations}
\begin{align}
    \frac{\Delta A_\mathrm{GWB}}{A_\mathrm{GWB}} &= 44 \% \times\, \left(\frac{\hat{\rho}}{25}\right)^{-2/5}, \label{eq:frac_A_uncert} \\
    \frac{\Delta \alpha}{\alpha} &= 40 \% \times\, \left(\frac{\hat{\rho}}{25}\right)^{-1/2}. \label{eq:frac_gam_uncert}
\end{align}
\label{eq:func_scaling_law}
\end{subequations}
The choice of using the lowest 5 frequencies to model the GWB does not affect the fractional parameter uncertainties until the signal has total S/N $\hat{\rho} \gtrsim 60$, whereupon using 30 frequencies improves their recovery. At such high detection significance, the GWB begins to dominate more than just the lowest 5 frequency bins, accruing greater significance and more informative constraints from higher frequencies. We also 
find that the inclusion of \textsc{BayesEphem} has minimal effect on the fractional uncertainty of the measured GWB parameters.

The measured spectral index of the GWB will be the primary arbiter of the source of the GWB. A spectral index of $\alpha = -2/3$, corresponding to our injected signal, is expected for a purely GW-driven population of inspiraling SMBHB systems, while other more exotic sources of the GWB are predicted to have different spectral indices.
At initial detection ($\hat{\rho} = 25$ or $\rho_{\rm HD} = 3$), the spectral index measurement should have a fractional uncertainty of 40\% (Eq.~\ref{eq:frac_gam_uncert}). In our suite of simulations, this precision is already sufficient to disfavor (at $95\%$ credibility) models of primordial GWs with matter \citep[$\alpha = -2$,][]{primordial_gw_spectrum} and radiation-dominated \citep[$\alpha = -1$,][]{lasky_pgw} equations of state.\footnote{But note that such a broadband primordial GWB signal with similar amplitude to the common-spectrum process found in the NANOGrav 12.5~yr analysis could be constrained at higher frequencies by ground-based GW interferometers and Big Bang nucleosynthesis \citep[e.g,][]{2021JCAP...01..071K}.} This precision is also sufficient to disfavor some models of the GWB produced by cosmic strings, such as those from kinks and cusps in the string loops which are predicted to have a spectral index, $\alpha = -7/6$ \citep{cosmic_string_spectrum}. As the significance of the GWB signal grows, the fractional uncertainty on the measured spectral index will continue to decrease and allow us to test other sources of the GWB. 
As we show later, an increase in the timing baseline of the dataset allows more comprehensive spectral modeling of the GWB, thereby reducing the need to rely on power-law fits.

\subsection{Milestone III: Properties Of The GWB} \label{subsec:gwb_spectrum}

Once there is sufficient evidence that the GWB is most likely due to a population of merging SMBHBs, we can use our spectral characterization to
probe the astrophysics of the underlying SMBHB population.
For example, the amplitude of a GWB produced by SMBHBs is set primarily by the mass distribution of SMBHs, often parametrized relative to their host galaxies \citep[e.g., in the M--$\sigma$ relation;][]{Gebhardt+2000, Ferrarese+Merritt-2000}, and the efficiency by which they reach sub-parsec separations.
The recovered amplitude can thus be used to distinguish between different population models of SMBHBs.  

Here we consider three such models \citep{McWilliams+2014,2016ApJ...826...11S,Sesana+201603}
with GWB amplitudes representative of typical values predicted in the literature: 
$\amp = \left[ 0.4^{+0.26}_{-0.16}, 1.5^{+0.3}_{-0.3}, 4.0^{+3.3}_{-1.8} \right] \times 10^{-15}$,
which were also examined in the NANOGrav 11 yr GWB analysis \citep{NG11_sgwb}.  
At initial detection, the amplitude measurement will have a fractional uncertainty of 44\%, which will be sufficient to distinguish the first model from the other two.
Models two and three are distinguishable with a fractional uncertainty of $\sim$37\% on the amplitude, occurring near the 17 yr slice of our simulated data.  
Thus, almost immediately after the initial detection of the GWB, we expect to be able to clearly distinguish between typical models in the literature.

\subsubsection{Constraining dynamical influences on the GWB spectral shape}

\begin{figure}[htb]
    \centering
    \includegraphics[width = \columnwidth]{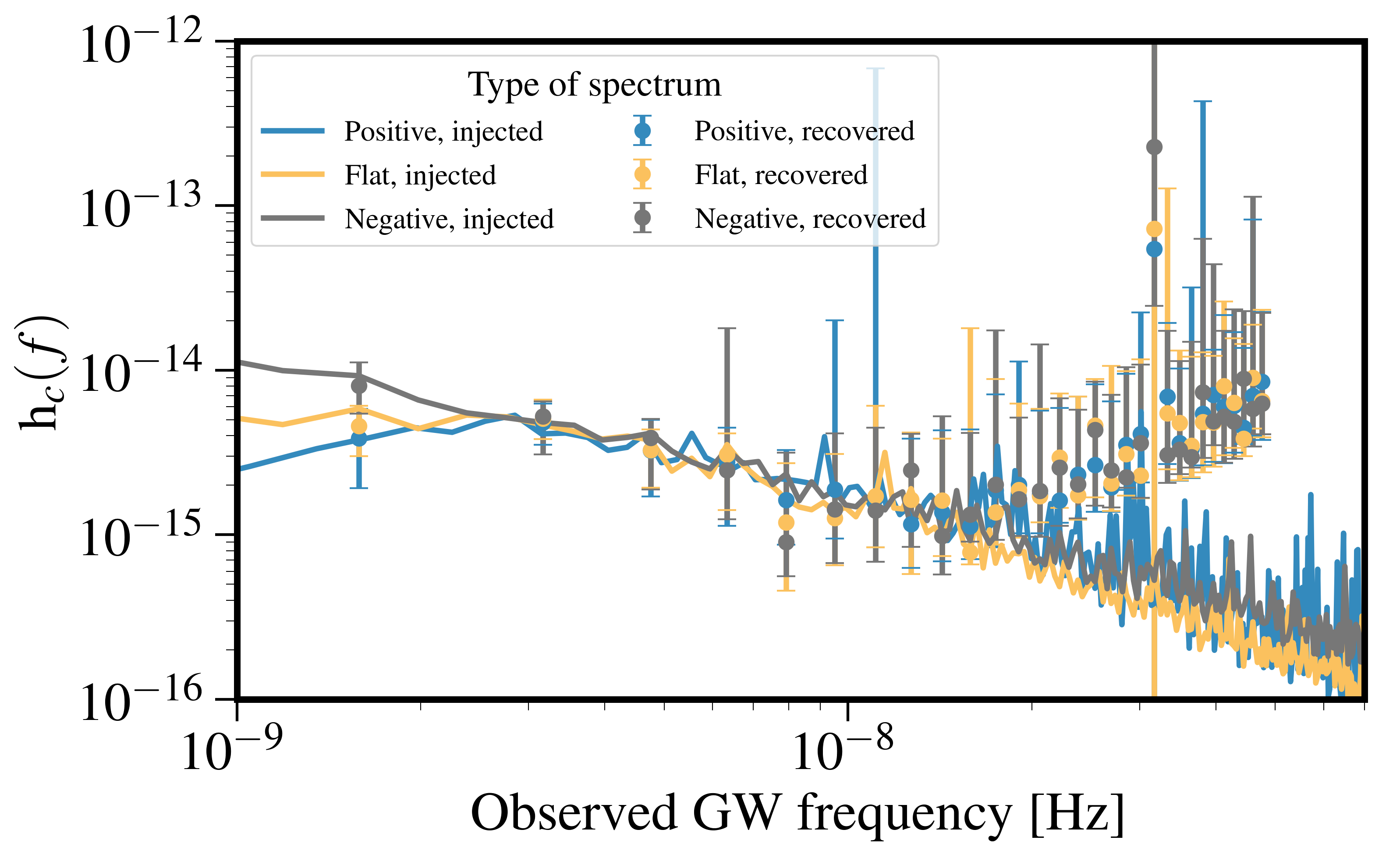}
    \caption{Per-frequency spectral modeling of the GWB spectrum for the 20 yr slice. Each frequency bin shows the median strain and uncertainty across ten realizations, while the injected GWB spectrum is shown by the solid lines. The lowest few frequency bins closely track the shape of the injected GWB, with the lowest frequency bin being the strongest discriminator between the three types of injected spectra. The high frequency bins are dominated by the white noise in the PTA and thus are insensitive to the GWB.}
    \label{fig:real_spectra_freespec_20yr}
\end{figure}

A realistic, astrophysical GWB from SMBHBs is, however, expected to deviate noticeably from the pure power-law spectrum we have assumed so far. To quantify the influence of these deviations on our detection prospects, we use astrophysically-motivated simulations of SMBHB populations to inject three non-power-law GWB spectra into the synthesized data.

To calculate plausible and self-consistent GWB spectra, we generate full mock populations of SMBH binaries.  We employ two entirely independent approaches to simulate SMBHB populations.  In the first, we anchor the population on the galaxy-galaxy merger rate derived from the Illustris cosmological simulations  \citep{Vogelsberger+2014a, Genel+2014, Vogelsberger+2014b}.  While the Illustris population of SMBH mergers has been used extensively \citep{Kelley+2017a, Kelley+2017b, Kelley+2018}, we have generalized our approach to allow for a more flexible range of binary properties, and thus resulting GWB spectra.  In particular, we have chosen binary hardening rates and masses so as to produce i) a GWB spectrum that is very nearly a power-law with a relatively steep ``negative''' spectral index at low frequencies (Fig.~\ref{fig:real_spectra_freespec_20yr}: grey), and ii) a strain spectrum that is nearly ``flat'' at low frequencies (Fig.~\ref{fig:real_spectra_freespec_20yr}: yellow).

The second method uses spectra originally generated for the astrophysical inference performed on the NANOGrav 11yr dataset \citep{NG11_sgwb}. This method uses a semi-analytic model \citep{2016ApJ...826...11S} to simulate a binary population using observational-based measurements of the galaxy stellar mass function, galaxy merger rate, and SMBH mass--host galaxy relation. The eccentricity distribution and binary hardening rate is incorporated over a wide range of parameters \citep{2017PhRvL.118r1102T}.  Model parameters are chosen to produce iii) a GW strain spectrum that is attenuated strongly enough to turn over, producing a ``positive'' spectral index at low frequencies (Fig.~\ref{fig:real_spectra_freespec_20yr}: blue).  

Overall, the amplitude of the simulated spectra is calibrated primarily by the distribution of SMBH masses, and the low-frequency spectral index is increased from the fiducial $-2/3$ by increasing the environmental hardening rate, mostly due to coupling with the nuclear stellar environment \citep{2016ApJ...826...11S, Kelley+2017b}.  For both methodologies, we have calibrated all three realistic spectra to have an amplitude of \gwbamp{1} at a frequency of 1 yr$^{-1}$, and are shown by the solid lines in Fig.~\ref{fig:real_spectra_freespec_20yr}. 

\begin{figure}[htb]
    \centering
    \includegraphics[width = \columnwidth]{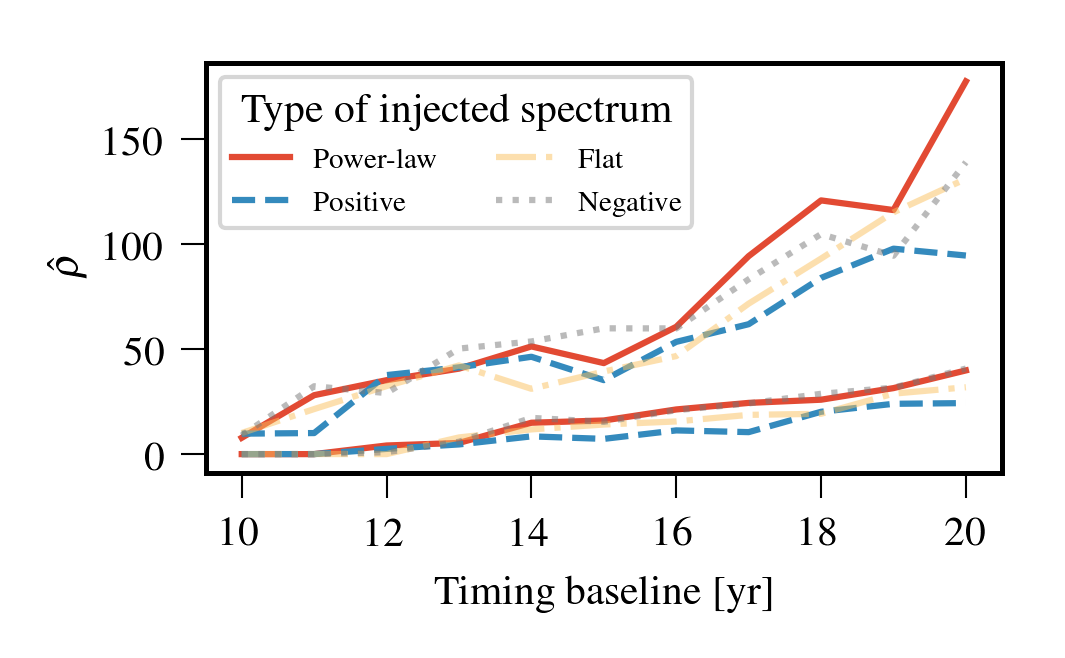}
    \caption{Detection significance for a variety of injected GWB spectra that are all modeled using a pure power-law spectrum in the detection pipeline. The two sets of curves represent the median 95\% credible interval upper and lower bounds on the S/N. We can see that power-law modeling of the GWB is robust for making a detection of the GWB.}
    \label{fig:real_spectra}
\end{figure}

The total S/N recovered using our standard detection pipeline (assuming a pure power-law strain spectrum) is shown for these realistically-modeled GWB spectra in Fig.~\ref{fig:real_spectra}. 
Even in the ``positive'' model, where a significant amount of GW strain is attenuated at low frequencies, the measured S/N is unaffected at the 15--17 yr slices (where initial detection is predicted to happen) and is only decreased by $\sim50\%$ at the 20 yr slice. This point is crucial, since it demonstrates that the assumption of a pure power-law strain spectrum allows for a high S/N detection, despite omitting the detailed spectral structure.  Additionally, even significant attenuation of the GWB signal at low frequencies does not represent a significant obstacle to detection. Any eventual reduction in the recovered S/N compared to the predicted S/N can then be used to interpret the presence of a turnover in the GWB spectrum.

However, many of the astrophysical details about SMBHB environments and demographics are missed by power-law analyses.  The additional encoded information can be extracted using per-frequency modeling of the GWB \citep{2013PhRvD..87j4021L,2013PhRvD..87d4035T}. In this approach, the strain in each frequency bin is modeled independently, allowing us to measure the spectral shape of the underlying GWB. A per-frequency spectral analysis of 20~yrs of data is shown in Fig.~\ref{fig:real_spectra_freespec_20yr} for our three astrophysically-motivated injected GWB spectra, with points and uncertainties representing strain measured independently in each frequency bin. We see that the lowest few bins are dominated by the injected GWB and closely track their spectral shapes. In a followup analysis we will explore full astrophysical model inference to determine how accurately astrophysical parameters can be extracted from realistic GWBs. However, the results shown here are very encouraging in that even models with some degeneracy (i.e., in the overall signal amplitude, $A_{\rm GWB}$) can be disentangled within the next several years.

\section{Accelerating PTA Milestones} \label{sec:ipta}

The S/N for a GWB signal in PTA data can be raised by increasing the time-span of the dataset, or by increasing the number of monitored pulsars \citep{os_scaling_laws_siemens}. However, new pulsars usually cannot be added into a PTA immediately upon discovery, and generally require multiple years of timing data to assess their appropriateness for PTA analysis. A solution is to combine the data already collected by PTAs around the world into a single dataset, and analyze this joint dataset for a GWB signal. Beyond NANOGrav, there are currently two other PTAs apart that have decades-worth of timing data on millisecond pulsars: the Parkes Pulsar Timing Array \citep[PPTA,][]{PPTA} and the European Pulsar Timing Array \citep[EPTA,][]{EPTA}. Together, these PTAs, along with emerging PTAs in India \citep[InPTA,][]{InPTA}, China \citep[CPTA,][]{CPTA} and South Africa \citep[SAPTA,][]{SKAPTA}, form the International Pulsar Timing Array \citep[IPTA,][]{IPTA}. To date, the IPTA has published two data releases \citep{ipta_dr1, ipta_dr2}, which result in an increase in the number of pulsars relative to individual regional PTAs at that time, while also increasing the time-span of data for pulsars that are common across the three PTAs. The IPTA has analyzed one of these datasets for GW signals \citep{ipta_dr1,ipta_dr1_gwb}, while the analysis of the latest data release \citep{ipta_dr2} is currently in progress. 

\begin{figure}[htb]
    \centering
    \includegraphics[width = \columnwidth]{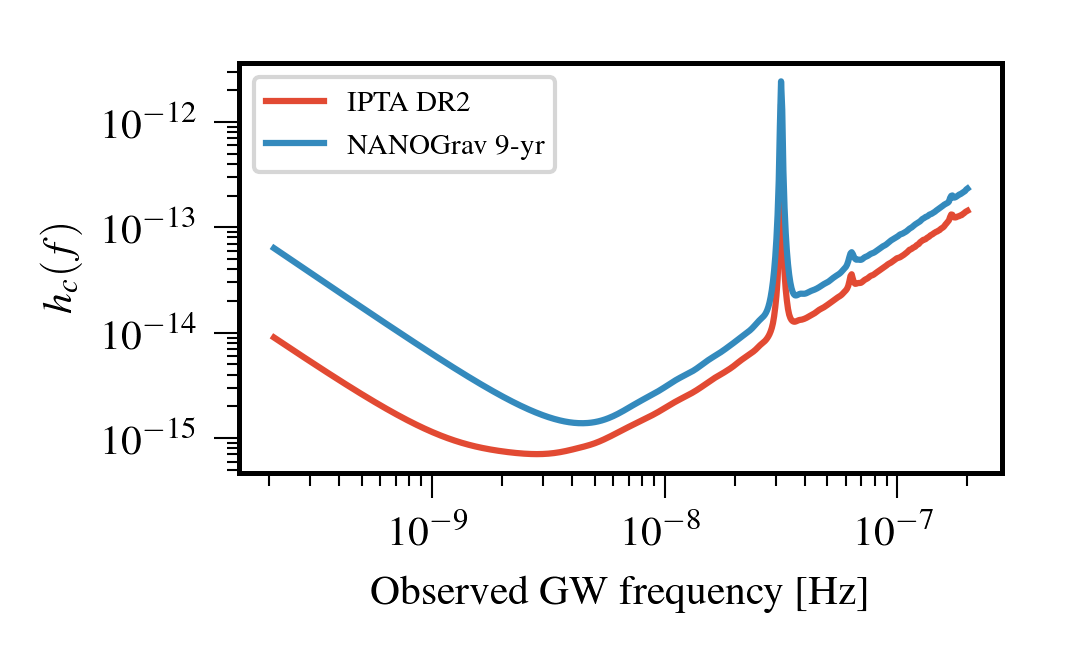}
    \caption{The sensitivity curve for IPTA dr2 and the NANOGrav 9 yr dataset, which was one of the components going into the construction of dr2. These sensitivity curves were constructed using {\sc hasasia}, and only used the white noise parameters released with IPTA dr2. The IPTA dr2 is more sensitive at all frequencies than the NANOGrav 9 yr dataset.}
    \label{fig:ipta_sc}
\end{figure}

To quantify the improvement offered by the IPTA over NANOGrav-only data, we use the IPTA's second data release \citep{ipta_dr2}, DR2, and compare its sensitivity to that of the NANOGrav subset of the DR2 dataset, i.e., the NANOGrav 9~yr dataset \citep{ng9_data} (NG9). 
We estimate the sensitivity using {\sc hasasia} \citep{hasasia_paper, hasasia_code}, a software package that can compute the per-frequency sensitivity of any PTA given the white and red noise properties of the pulsars. We use the noise properties of the pulsars provided with the public DR2 data, but do not include intrinsic red noise since these parameters are highly covariant with a common process and would require a full GWB detection analysis for accurate calculation.

The per-frequency sensitivities of IPTA DR2 and NG9 are shown in Fig.~\ref{fig:ipta_sc}. IPTA DR2 is more sensitive across all frequencies than NG9. 
Consequently, IPTA DR2 achieves a total-signal significance that is a factor of 30 higher than NG9. This corresponds to a decrease in the fractional uncertainty on the the GWB amplitude and spectral index by factors of 4 and 5.5 respectively. Note that these are only estimates since (as previously mentioned) no intrinsic red noise was accounted for in the sensitivity curves, which would have the effect of reducing each of the S/N values. However, given these estimates, an analysis of a combined IPTA dataset offers the opportunity to detect a GWB and extract the underlying astrophysics of the source population at a faster pace and with higher significance than can be achieved with any individual PTA.

\section{Conclusion} \label{sec:conclusion}

The recent NANOGrav evidence of a common-spectrum process is a tantalizing hint of the first signs of a GWB with growing significance in PTA data. If so, we have developed a road-map for the next several years that includes three key PTA scientific milestones: (I) robust detection of the GWB through measuring Hellings-\&-Downs-correlated timing delays across the PTA; (II) discriminating the origin of the GWB as being from either a population of supermassive black-hole binaries or exotica; and (III) unveiling the demographic and dynamical properties of supermassive black-hole binaries as encoded in the shape of the recovered GWB strain spectrum. All of these milestones could be achievable by the middle of this decade under conservative assumptions, and even sooner under the auspices of the 
IPTA. 

\acknowledgements

This work has been carried out by the NANOGrav collaboration, which is part of the International Pulsar Timing Array. 
The NANOGrav project receives support from National Science Foundation (NSF) Physics Frontiers Center award number 1430284.
This work made use of the Super Computing System (Spruce Knob) at West Virginia University (WVU), which are funded in part by the National Science Foundation EPSCoR Research Infrastructure Improvement Cooperative Agreement \#1003907, the state of West Virginia (WVEPSCoR via the Higher Education Policy Commission) and WVU.  Some of our models utilized the IllustrisTNG simulations public data release \citep[\url{http://www.tng-project.org/};][]{Nelson+Springel+2019}, and \texttt{sublink} merger trees \citep{Rodriguez-Gomez+2015}.

The Flatiron Institute is supported by the Simons Foundation. 
Pulsar research at UBC is supported by an NSERC Discovery Grant and by the Canadian Institute for Advanced Research. 
JS and MV acknowledge support from the JPL RTD program. 
SBS acknowledges support for this work from NSF grants \#1458952 and \#1815664. SBS is a CIFAR Azrieli Global Scholar in the Gravity and the Extreme Universe program. SRT acknowledges support from NSF grant AST-\#2007993, and a Dean's Faculty Fellowship from Vanderbilt Unviversity's College of Arts \& Science.
TTP acknowledges support from the MTA-ELTE Extragalactic Astrophysics Research Group, funded by the Hungarian Academy of Sciences (Magyar Tudom\'{a}nyos Akad\'{e}mia), that was used during the development of this research.
TD and ML acknowledge NSF AAG award number 2009468.

\software{\texttt{ENTERPRISE} \citep{enterprise}, \texttt{enterprise\_extensions} \citep{enterprise_ext}, \texttt{hasasia} \citep{hasasia}, \texttt{libstempo} \citep{libstempo}, \texttt{matplotlib} \citep{matplotlib}, \texttt{PTMCMC} \citep{ptmcmc}}

\bibliography{bibliography.bib}

\begin{thebibliography}{}
\expandafter\ifx\csname natexlab\endcsname\relax\def\natexlab#1{#1}\fi
\providecommand{\url}[1]{\href{#1}{#1}}
\providecommand{\dodoi}[1]{doi:~\href{http://doi.org/#1}{\nolinkurl{#1}}}
\providecommand{\doeprint}[1]{\href{http://ascl.net/#1}{\nolinkurl{http://ascl.net/#1}}}
\providecommand{\doarXiv}[1]{\href{https://arxiv.org/abs/#1}{\nolinkurl{https://arxiv.org/abs/#1}}}

\bibitem[{{Accadia} {et~al.}(2012){Accadia}, {Acernese}, {Alshourbagy},
  {Amico}, {Antonucci}, {Aoudia}, {Arnaud}, {Arnault}, {Arun}, {Astone}, \&
  et~al.}]{VIRGO_detector_ref}
{Accadia}, T., {Acernese}, F., {Alshourbagy}, M., {et~al.} 2012, Journal of
  Instrumentation, 7, 3012, \dodoi{10.1088/1748-0221/7/03/P03012}

\bibitem[{{Alam} {et~al.}(2021){Alam}, {Arzoumanian}, {Baker}, {Blumer},
  {Bohler}, {Brazier}, {Brook}, {Burke-Spolaor}, {Caballero}, {Camuccio},
  {Chamberlain}, {Chatterjee}, {Cordes}, {Cornish}, {Crawford}, {Cromartie},
  {Decesar}, {Demorest}, {Dolch}, {Ellis}, {Ferdman}, {Ferrara}, {Fiore},
  {Fonseca}, {Garcia}, {Garver-Daniels}, {Gentile}, {Good}, {Gusdorff},
  {Halmrast}, {Hazboun}, {Islo}, {Jennings}, {Jessup}, {Jones}, {Kaiser},
  {Kaplan}, {Kelley}, {Key}, {Lam}, {Lazio}, {Lorimer}, {Luo}, {Lynch},
  {Madison}, {Maraccini}, {McLaughlin}, {Mingarelli}, {Ng}, {Nguyen}, {Nice},
  {Pennucci}, {Pol}, {Ramette}, {Ransom}, {Ray}, {Shapiro-Albert}, {Siemens},
  {Simon}, {Spiewak}, {Stairs}, {Stinebring}, {Stovall}, {Swiggum}, {Taylor},
  {Tripepi}, {Vallisneri}, {Vigeland}, {Witt}, {Zhu}, \& {Nanograv
  Collaboration}}]{NG12p5_timing}
{Alam}, M.~F., {Arzoumanian}, Z., {Baker}, P.~T., {et~al.} 2021, \apjs, 252, 4,
  \dodoi{10.3847/1538-4365/abc6a0}

\bibitem[{{Anholm} {et~al.}(2009){Anholm}, {Ballmer}, {Creighton}, {Price}, \&
  {Siemens}}]{abc+2009}
{Anholm}, M., {Ballmer}, S., {Creighton}, J.~D.~E., {Price}, L.~R., \&
  {Siemens}, X. 2009, \prd, 79, 084030, \dodoi{10.1103/PhysRevD.79.084030}

\bibitem[{{Arzoumanian} {et~al.}(2016){Arzoumanian}, {Brazier},
  {Burke-Spolaor}, {Chamberlin}, {Chatterjee}, {Christy}, {Cordes}, {Cornish},
  {Crowter}, {Demorest}, {Deng}, {Dolch}, {Ellis}, {Ferdman}, {Fonseca},
  {Garver-Daniels}, {Gonzalez}, {Jenet}, {Jones}, {Jones}, {Kaspi}, {Koop},
  {Lam}, {Lazio}, {Levin}, {Lommen}, {Lorimer}, {Luo}, {Lynch}, {Madison},
  {McLaughlin}, {McWilliams}, {Mingarelli}, {Nice}, {Palliyaguru}, {Pennucci},
  {Ransom}, {Sampson}, {Sanidas}, {Sesana}, {Siemens}, {Simon}, {Stairs},
  {Stinebring}, {Stovall}, {Swiggum}, {Taylor}, {Vallisneri}, {van Haasteren},
  {Wang}, {Zhu}, \& {NANOGrav Collaboration}}]{NG9_sgwb}
{Arzoumanian}, Z., {Brazier}, A., {Burke-Spolaor}, S., {et~al.} 2016, \apj,
  821, 13, \dodoi{10.3847/0004-637X/821/1/13}

\bibitem[{{Arzoumanian} {et~al.}(2018){Arzoumanian}, {Baker}, {Brazier},
  {Burke-Spolaor}, {Chamberlin}, {Chatterjee}, {Christy}, {Cordes}, {Cornish},
  {Crawford}, {Thankful Cromartie}, {Crowter}, {DeCesar}, {Demorest}, {Dolch},
  {Ellis}, {Ferdman}, {Ferrara}, {Folkner}, {Fonseca}, {Garver-Daniels},
  {Gentile}, {Haas}, {Hazboun}, {Huerta}, {Islo}, {Jones}, {Jones}, {Kaplan},
  {Kaspi}, {Lam}, {Lazio}, {Levin}, {Lommen}, {Lorimer}, {Luo}, {Lynch},
  {Madison}, {McLaughlin}, {McWilliams}, {Mingarelli}, {Ng}, {Nice}, {Park},
  {Pennucci}, {Pol}, {Ransom}, {Ray}, {Rasskazov}, {Siemens}, {Simon},
  {Spiewak}, {Stairs}, {Stinebring}, {Stovall}, {Swiggum}, {Taylor},
  {Vallisneri}, {van Haasteren}, {Vigeland }, {Zhu}, \& {NANOGrav
  Collaboration}}]{NG11_sgwb}
{Arzoumanian}, Z., {Baker}, P.~T., {Brazier}, A., {et~al.} 2018, \apj, 859, 47,
  \dodoi{10.3847/1538-4357/aabd3b}

\bibitem[{{Arzoumanian} {et~al.}(2020){Arzoumanian}, {Baker}, {Blumer},
  {B{\'e}csy}, {Brazier}, {Brook}, {Burke-Spolaor}, {Chatterjee}, {Chen},
  {Cordes}, {Cornish}, {Crawford}, {Cromartie}, {Decesar}, {Demorest}, {Dolch},
  {Ellis}, {Ferrara}, {Fiore}, {Fonseca}, {Garver-Daniels}, {Gentile}, {Good},
  {Hazboun}, {Holgado}, {Islo}, {Jennings}, {Jones}, {Kaiser}, {Kaplan},
  {Kelley}, {Key}, {Laal}, {Lam}, {Lazio}, {Lorimer}, {Luo}, {Lynch},
  {Madison}, {McLaughlin}, {Mingarelli}, {Ng}, {Nice}, {Pennucci}, {Pol},
  {Ransom}, {Ray}, {Shapiro-Albert}, {Siemens}, {Simon}, {Spiewak}, {Stairs},
  {Stinebring}, {Stovall}, {Sun}, {Swiggum}, {Taylor}, {Turner}, {Vallisneri},
  {Vigeland}, {Witt}, \& {Nanograv Collaboration}}]{NG12p5_gwb}
{Arzoumanian}, Z., {Baker}, P.~T., {Blumer}, H., {et~al.} 2020, \apjl, 905,
  L34, \dodoi{10.3847/2041-8213/abd401}

\bibitem[{{Bailes} {et~al.}(2016){Bailes}, {Barr}, {Bhat}, {Brink}, {Buchner},
  {Burgay}, {Camilo}, {Champion}, {Hessels}, {Jameson}, {Johnston},
  {Karastergiou}, {Karuppusamy}, {Kaspi}, {Keith}, {Kramer}, {McLaughlin},
  {Moodley}, {Oslowski}, {Possenti}, {Ransom}, {Rasio}, {Sievers}, {Serylak},
  {Stappers}, {Stairs}, {Theureau}, {van Straten}, {Weltevrede}, \&
  {Wex}}]{SKAPTA}
{Bailes}, M., {Barr}, E., {Bhat}, N.~D.~R., {et~al.} 2016, in MeerKAT Science:
  On the Pathway to the SKA, 11.
\newblock \doarXiv{1803.07424}

\bibitem[{{Begelman} {et~al.}(1980){Begelman}, {Blandford}, \&
  {Rees}}]{Begelman+1980}
{Begelman}, M.~C., {Blandford}, R.~D., \& {Rees}, M.~J. 1980, nat, 287, 307,
  \dodoi{10.1038/287307a0}

\bibitem[{Blasi {et~al.}(2021)Blasi, Brdar, \& Schmitz}]{ng12p5_cs_2}
Blasi, S., Brdar, V., \& Schmitz, K. 2021, Phys. Rev. Lett., 126, 041305,
  \dodoi{10.1103/PhysRevLett.126.041305}

\bibitem[{{Burke-Spolaor} {et~al.}(2019){Burke-Spolaor}, {Taylor}, {Charisi},
  {Dolch}, {Hazboun}, {Holgado}, {Kelley}, {Lazio}, {Madison}, {McMann},
  {Mingarelli}, {Rasskazov}, {Siemens}, {Simon}, \&
  {Smith}}]{2019A&ARv..27....5B}
{Burke-Spolaor}, S., {Taylor}, S.~R., {Charisi}, M., {et~al.} 2019, \aapr, 27,
  5, \dodoi{10.1007/s00159-019-0115-7}

\bibitem[{{Chamberlin} {et~al.}(2015){Chamberlin}, {Creighton}, {Siemens},
  {Demorest}, {Ellis}, {Price}, \& {Romano}}]{ccs+2015}
{Chamberlin}, S.~J., {Creighton}, J.~D.~E., {Siemens}, X., {et~al.} 2015, \prd,
  91, 044048, \dodoi{10.1103/PhysRevD.91.044048}

\bibitem[{{Charisi} {et~al.}(2016){Charisi}, {Bartos}, {Haiman},
  {Price-Whelan}, {Graham}, {Bellm}, {Laher}, \&
  {M{\'a}rka}}]{2016MNRAS.463.2145C}
{Charisi}, M., {Bartos}, I., {Haiman}, Z., {et~al.} 2016, \mnras, 463, 2145,
  \dodoi{10.1093/mnras/stw1838}

\bibitem[{{Chen} {et~al.}(2020){Chen}, {Liu}, {Liao}, {Holgado}, {Guo},
  {Gruendl}, {Morganson}, {Shen}, {Zhang}, {Abbott}, {Aguena}, {Allam},
  {Avila}, {Bertin}, {Bhargava}, {Brooks}, {Burke}, {Carnero Rosell},
  {Carollo}, {Carrasco Kind}, {Carretero}, {Costanzi}, {da Costa}, {Davis}, {De
  Vicente}, {Desai}, {Diehl}, {Doel}, {Everett}, {Flaugher}, {Friedel},
  {Frieman}, {Garc{\'\i}a-Bellido}, {Gaztanaga}, {Glazebrook}, {Gruen},
  {Gutierrez}, {Hinton}, {Hollowood}, {James}, {Kim}, {Kuehn}, {Kuropatkin},
  {Lewis}, {Lidman}, {Lima}, {Maia}, {March}, {Marshall}, {Menanteau},
  {Miquel}, {Palmese}, {Paz-Chinch{\'o}n}, {Plazas}, {Sanchez}, {Schubnell},
  {Serrano}, {Sevilla-Noarbe}, {Smith}, {Suchyta}, {Swanson}, {Tarle},
  {Tucker}, {Varga}, \& {Walker}}]{2020arXiv200812329C}
{Chen}, Y.-C., {Liu}, X., {Liao}, W.-T., {et~al.} 2020, arXiv e-prints,
  arXiv:2008.12329.
\newblock \doarXiv{2008.12329}

\bibitem[{De~Luca {et~al.}(2021)De~Luca, Franciolini, \& Riotto}]{ng12p5_pbh_2}
De~Luca, V., Franciolini, G., \& Riotto, A. 2021, Phys. Rev. Lett., 126,
  041303, \dodoi{10.1103/PhysRevLett.126.041303}

\bibitem[{{Demorest} {et~al.}(2013){Demorest}, {Ferdman}, {Gonzalez}, {Nice},
  {Ransom}, {Stairs}, {Arzoumanian}, {Brazier}, {Burke-Spolaor}, {Chamberlin},
  {Cordes}, {Ellis}, {Finn}, {Freire}, {Giampanis}, {Jenet}, {Kaspi}, {Lazio},
  {Lommen}, {McLaughlin}, {Palliyaguru}, {Perrodin}, {Shannon}, {Siemens},
  {Stinebring}, {Swiggum}, \& {Zhu}}]{dfg+13}
{Demorest}, P.~B., {Ferdman}, R.~D., {Gonzalez}, M.~E., {et~al.} 2013, \apj,
  762, 94, \dodoi{10.1088/0004-637X/762/2/94}

\bibitem[{{Desvignes} {et~al.}(2016){Desvignes}, {Caballero}, {Lentati},
  {Verbiest}, {Champion}, {Stappers}, {Janssen}, {Lazarus}, {Os{\l}owski},
  {Babak}, {Bassa}, {Brem}, {Burgay}, {Cognard}, {Gair}, {Graikou},
  {Guillemot}, {Hessels}, {Jessner}, {Jordan}, {Karuppusamy}, {Kramer},
  {Lassus}, {Lazaridis}, {Lee}, {Liu}, {Lyne}, {McKee}, {Mingarelli},
  {Perrodin}, {Petiteau}, {Possenti}, {Purver}, {Rosado}, {Sanidas}, {Sesana},
  {Shaifullah}, {Smits}, {Taylor}, {Theureau}, {Tiburzi}, {van Haasteren}, \&
  {Vecchio}}]{EPTA_dr1}
{Desvignes}, G., {Caballero}, R.~N., {Lentati}, L., {et~al.} 2016, \mnras, 458,
  3341, \dodoi{10.1093/mnras/stw483}

\bibitem[{{Detweiler}(1979)}]{PTA_detweiler}
{Detweiler}, S. 1979, \apj, 234, 1100, \dodoi{10.1086/157593}

\bibitem[{Ellis \& Lewicki(2021)}]{ng12p5_cs_1}
Ellis, J., \& Lewicki, M. 2021, Phys. Rev. Lett., 126, 041304,
  \dodoi{10.1103/PhysRevLett.126.041304}

\bibitem[{Ellis \& van Haasteren(2017)}]{ptmcmc}
Ellis, J., \& van Haasteren, R. 2017, jellis18/PTMCMCSampler: Official Release,
  \dodoi{10.5281/zenodo.1037579}

\bibitem[{{Ellis} {et~al.}(2019){Ellis}, {Vallisneri}, {Taylor}, \&
  {Baker}}]{enterprise}
{Ellis}, J.~A., {Vallisneri}, M., {Taylor}, S.~R., \& {Baker}, P.~T. 2019,
  {ENTERPRISE: Enhanced Numerical Toolbox Enabling a Robust PulsaR Inference
  SuitE}.
\newblock \doeprint{1912.015}

\bibitem[{{Ferrarese} \& {Merritt}(2000)}]{Ferrarese+Merritt-2000}
{Ferrarese}, L., \& {Merritt}, D. 2000, \apjl, 539, L9, \dodoi{10.1086/312838}

\bibitem[{Folkner \& Park(2018)}]{de438}
Folkner, W.~M., \& Park, R.~S. 2018, Planetary ephemeris DE438 for Juno, Tech.
  Rep. IOM 392R-18-004, Jet Propulsion Laboratory, Pasadena, CA

\bibitem[{{Foster} \& {Backer}(1990)}]{PTA_foster_backer}
{Foster}, R.~S., \& {Backer}, D.~C. 1990, \apj, 361, 300,
  \dodoi{10.1086/169195}

\bibitem[{{Gebhardt} {et~al.}(2000){Gebhardt}, {Bender}, {Bower}, {Dressler},
  {Faber}, {Filippenko}, {Green}, {Grillmair}, {Ho}, {Kormendy}, {Lauer},
  {Magorrian}, {Pinkney}, {Richstone}, \& {Tremaine}}]{Gebhardt+2000}
{Gebhardt}, K., {Bender}, R., {Bower}, G., {et~al.} 2000, \apjl, 539, L13,
  \dodoi{10.1086/312840}

\bibitem[{{Genel} {et~al.}(2014){Genel}, {Vogelsberger}, {Springel}, {Sijacki},
  {Nelson}, {Snyder}, {Rodriguez-Gomez}, {Torrey}, \& {Hernquist}}]{Genel+2014}
{Genel}, S., {Vogelsberger}, M., {Springel}, V., {et~al.} 2014, \mnras, 445,
  175, \dodoi{10.1093/mnras/stu1654}

\bibitem[{{Graham} {et~al.}(2015){Graham}, {Djorgovski}, {Stern}, {Drake},
  {Mahabal}, {Donalek}, {Glikman}, {Larson}, \&
  {Christensen}}]{2015MNRAS.453.1562G}
{Graham}, M.~J., {Djorgovski}, S.~G., {Stern}, D., {et~al.} 2015, \mnras, 453,
  1562, \dodoi{10.1093/mnras/stv1726}

\bibitem[{{Grishchuk}(2005)}]{primordial_gw_spectrum}
{Grishchuk}, L.~P. 2005, Physics Uspekhi, 48, 1235,
  \dodoi{10.1070/PU2005v048n12ABEH005795}

\bibitem[{{Harry} \& {LIGO Scientific Collaboration}(2010)}]{LIGO_detector_ref}
{Harry}, G.~M., \& {LIGO Scientific Collaboration}. 2010, Classical and Quantum
  Gravity, 27, 084006, \dodoi{10.1088/0264-9381/27/8/084006}

\bibitem[{{Hazboun} {et~al.}(2019{\natexlab{a}}){Hazboun}, {Romano}, \&
  {Smith}}]{hasasia_code}
{Hazboun}, J., {Romano}, J., \& {Smith}, T. 2019{\natexlab{a}}, The Journal of
  Open Source Software, 4, 1775, \dodoi{10.21105/joss.01775}

\bibitem[{{Hazboun} {et~al.}(2019{\natexlab{b}}){Hazboun}, {Romano}, \&
  {Smith}}]{hasasia}
---. 2019{\natexlab{b}}, The Journal of Open Source Software, 4, 1775,
  \dodoi{10.21105/joss.01775}

\bibitem[{{Hazboun} {et~al.}(2019{\natexlab{c}}){Hazboun}, {Romano}, \&
  {Smith}}]{hasasia_paper}
{Hazboun}, J.~S., {Romano}, J.~D., \& {Smith}, T.~L. 2019{\natexlab{c}}, \prd,
  100, 104028, \dodoi{10.1103/PhysRevD.100.104028}

\bibitem[{{Hellings} \& {Downs}(1983)}]{HD_curve}
{Hellings}, R.~W., \& {Downs}, G.~S. 1983, \apjl, 265, L39,
  \dodoi{10.1086/183954}

\bibitem[{{Hobbs}(2013)}]{PPTA}
{Hobbs}, G. 2013, Classical and Quantum Gravity, 30, 224007,
  \dodoi{10.1088/0264-9381/30/22/224007}

\bibitem[{{Hobbs} {et~al.}(2010){Hobbs}, {Archibald}, {Arzoumanian}, {Backer},
  {Bailes}, {Bhat}, {Burgay}, {Burke-Spolaor}, {Champion}, {Cognard}, {Coles},
  {Cordes}, {Demorest}, {Desvignes}, {Ferdman}, {Finn}, {Freire}, {Gonzalez},
  {Hessels}, {Hotan}, {Janssen}, {Jenet}, {Jessner}, {Jordan}, {Kaspi},
  {Kramer}, {Kondratiev}, {Lazio}, {Lazaridis}, {Lee}, {Levin}, {Lommen},
  {Lorimer}, {Lynch}, {Lyne}, {Manchester}, {McLaughlin}, {Nice}, {Oslowski},
  {Pilia}, {Possenti}, {Purver}, {Ransom}, {Reynolds}, {Sanidas}, {Sarkissian},
  {Sesana}, {Shannon}, {Siemens}, {Stairs}, {Stappers}, {Stinebring},
  {Theureau}, {van Haasteren}, {van Straten}, {Verbiest}, {Yardley}, \&
  {You}}]{IPTA}
{Hobbs}, G., {Archibald}, A., {Arzoumanian}, Z., {et~al.} 2010, Classical and
  Quantum Gravity, 27, 084013, \dodoi{10.1088/0264-9381/27/8/084013}

\bibitem[{{Hunter}(2007)}]{matplotlib}
{Hunter}, J.~D. 2007, Computing in Science and Engineering, 9, 90,
  \dodoi{10.1109/MCSE.2007.55}

\bibitem[{{Jaffe} \& {Backer}(2003)}]{2003ApJ...583..616J}
{Jaffe}, A.~H., \& {Backer}, D.~C. 2003, \apj, 583, 616, \dodoi{10.1086/345443}

\bibitem[{{Joshi} {et~al.}(2018){Joshi}, {Arumugasamy}, {Bagchi},
  {Bandyopadhyay}, {Basu}, {Dhand a Batra}, {Bethapudi}, {Choudhary}, {De},
  {Dey}, {Gopakumar}, {Gupta}, {Krishnakumar}, {Maan}, {Manoharan}, {Naidu},
  {Nandi}, {Pathak}, {Surnis}, \& {Susobhanan}}]{InPTA}
{Joshi}, B.~C., {Arumugasamy}, P., {Bagchi}, M., {et~al.} 2018, Journal of
  Astrophysics and Astronomy, 39, 51, \dodoi{10.1007/s12036-018-9549-y}

\bibitem[{{Kelley} {et~al.}(2017{\natexlab{a}}){Kelley}, {Blecha}, \&
  {Hernquist}}]{Kelley+2017a}
{Kelley}, L.~Z., {Blecha}, L., \& {Hernquist}, L. 2017{\natexlab{a}}, \mnras,
  464, 3131, \dodoi{10.1093/mnras/stw2452}

\bibitem[{{Kelley} {et~al.}(2017{\natexlab{b}}){Kelley}, {Blecha}, {Hernquist},
  {Sesana}, \& {Taylor}}]{2017MNRAS.471.4508K}
{Kelley}, L.~Z., {Blecha}, L., {Hernquist}, L., {Sesana}, A., \& {Taylor},
  S.~R. 2017{\natexlab{b}}, \mnras, 471, 4508, \dodoi{10.1093/mnras/stx1638}

\bibitem[{{Kelley} {et~al.}(2017{\natexlab{c}}){Kelley}, {Blecha}, {Hernquist},
  {Sesana}, \& {Taylor}}]{Kelley+2017b}
---. 2017{\natexlab{c}}, \mnras, 471, 4508, \dodoi{10.1093/mnras/stx1638}

\bibitem[{{Kelley} {et~al.}(2018){Kelley}, {Blecha}, {Hernquist}, {Sesana}, \&
  {Taylor}}]{Kelley+2018}
---. 2018, \mnras, 477, 964, \dodoi{10.1093/mnras/sty689}

\bibitem[{{Kerr} {et~al.}(2020){Kerr}, {Reardon}, {Hobbs}, {Shannon},
  {Manchester}, {Dai}, {Russell}, {Zhang}, {van Straten}, {Os{\l}owski},
  {Parthasarathy}, {Spiewak}, {Bailes}, {Bhat}, {Cameron}, {Coles}, {Dempsey},
  {Deng}, {Goncharov}, {Kaczmarek}, {Keith}, {Lasky}, {Lower}, {Preisig},
  {Sarkissian}, {Toomey}, {Wang}, {Wang}, {Zhang}, \& {Zhu}}]{PPTA_dr2}
{Kerr}, M., {Reardon}, D.~J., {Hobbs}, G., {et~al.} 2020, \pasa, 37, e020,
  \dodoi{10.1017/pasa.2020.11}

\bibitem[{{Kramer} \& {Champion}(2013)}]{EPTA}
{Kramer}, M., \& {Champion}, D.~J. 2013, Classical and Quantum Gravity, 30,
  224009, \dodoi{10.1088/0264-9381/30/22/224009}

\bibitem[{{Kuroyanagi} {et~al.}(2021){Kuroyanagi}, {Takahashi}, \&
  {Yokoyama}}]{2021JCAP...01..071K}
{Kuroyanagi}, S., {Takahashi}, T., \& {Yokoyama}, S. 2021, \jcap, 2021, 071,
  \dodoi{10.1088/1475-7516/2021/01/071}

\bibitem[{{Lasky} {et~al.}(2016){Lasky}, {Mingarelli}, {Smith}, {Giblin},
  {Thrane}, {Reardon}, {Caldwell}, {Bailes}, {Bhat}, {Burke-Spolaor}, {Dai},
  {Dempsey}, {Hobbs}, {Kerr}, {Levin}, {Manchester}, {Os{\l}owski}, {Ravi},
  {Rosado}, {Shannon}, {Spiewak}, {van Straten}, {Toomey}, {Wang}, {Wen},
  {You}, \& {Zhu}}]{lasky_pgw}
{Lasky}, P.~D., {Mingarelli}, C. M.~F., {Smith}, T.~L., {et~al.} 2016, Physical
  Review X, 6, 011035, \dodoi{10.1103/PhysRevX.6.011035}

\bibitem[{{Lee}(2016)}]{CPTA}
{Lee}, K.~J. 2016, in Astronomical Society of the Pacific Conference Series,
  Vol. 502, Frontiers in Radio Astronomy and FAST Early Sciences Symposium
  2015, ed. L.~{Qain} \& D.~{Li}, 19

\bibitem[{{Lentati} {et~al.}(2013){Lentati}, {Alexander}, {Hobson}, {Taylor},
  {Gair}, {Balan}, \& {van Haasteren}}]{2013PhRvD..87j4021L}
{Lentati}, L., {Alexander}, P., {Hobson}, M.~P., {et~al.} 2013, \prd, 87,
  104021, \dodoi{10.1103/PhysRevD.87.104021}

\bibitem[{{Lentati} {et~al.}(2016){Lentati}, {Shannon}, {Coles}, {Verbiest},
  {van Haasteren}, {Ellis}, {Caballero}, {Manchester}, {Arzoumanian}, {Babak},
  {Bassa}, {Bhat}, {Brem}, {Burgay}, {Burke-Spolaor}, {Champion}, {Chatterjee},
  {Cognard}, {Cordes}, {Dai}, {Demorest}, {Desvignes}, {Dolch}, {Ferdman},
  {Fonseca}, {Gair}, {Gonzalez}, {Graikou}, {Guillemot}, {Hessels}, {Hobbs},
  {Janssen}, {Jones}, {Karuppusamy}, {Keith}, {Kerr}, {Kramer}, {Lam}, {Lasky},
  {Lassus}, {Lazarus}, {Lazio}, {Lee}, {Levin}, {Liu}, {Lynch}, {Madison},
  {McKee}, {McLaughlin}, {McWilliams}, {Mingarelli}, {Nice}, {Os{\l}owski},
  {Pennucci}, {Perera}, {Perrodin}, {Petiteau}, {Possenti}, {Ransom},
  {Reardon}, {Rosado}, {Sanidas}, {Sesana}, {Shaifullah}, {Siemens}, {Smits},
  {Stairs}, {Stappers}, {Stinebring}, {Stovall}, {Swiggum}, {Taylor},
  {Theureau}, {Tiburzi}, {Toomey}, {Vallisneri}, {van Straten}, {Vecchio},
  {Wang}, {Wang}, {You}, {Zhu}, \& {Zhu}}]{ipta_dr1_gwb}
{Lentati}, L., {Shannon}, R.~M., {Coles}, W.~A., {et~al.} 2016, \mnras, 458,
  2161, \dodoi{10.1093/mnras/stw395}

\bibitem[{{Liu} {et~al.}(2019){Liu}, {Gezari}, {Ayers}, {Burgett}, {Chambers},
  {Hodapp}, {Huber}, {Kudritzki}, {Metcalfe}, {Tonry}, {Wainscoat}, \&
  {Waters}}]{2019ApJ...884...36L}
{Liu}, T., {Gezari}, S., {Ayers}, M., {et~al.} 2019, \apj, 884, 36,
  \dodoi{10.3847/1538-4357/ab40cb}

\bibitem[{{McWilliams} {et~al.}(2014){McWilliams}, {Ostriker}, \&
  {Pretorius}}]{McWilliams+2014}
{McWilliams}, S.~T., {Ostriker}, J.~P., \& {Pretorius}, F. 2014, \apj, 789,
  156, \dodoi{10.1088/0004-637X/789/2/156}

\bibitem[{{NANOGrav Collaboration} {et~al.}(2015){NANOGrav Collaboration},
  {Arzoumanian}, {Brazier}, {Burke-Spolaor}, {Chamberlin}, {Chatterjee},
  {Christy}, {Cordes}, {Cornish}, {Crowter}, {Demorest}, {Dolch}, {Ellis},
  {Ferdman}, {Fonseca}, {Garver-Daniels}, {Gonzalez}, {Jenet}, {Jones},
  {Jones}, {Kaspi}, {Koop}, {Lam}, {Lazio}, {Levin}, {Lommen}, {Lorimer},
  {Luo}, {Lynch}, {Madison}, {McLaughlin}, {McWilliams}, {Nice}, {Palliyaguru},
  {Pennucci}, {Ransom}, {Siemens}, {Stairs}, {Stinebring}, {Stovall},
  {Swiggum}, {Vallisneri}, {van Haasteren}, {Wang}, \& {Zhu}}]{ng9_data}
{NANOGrav Collaboration}, {Arzoumanian}, Z., {Brazier}, A., {et~al.} 2015,
  \apj, 813, 65, \dodoi{10.1088/0004-637X/813/1/65}

\bibitem[{{Nelson} {et~al.}(2019){Nelson}, {Springel}, {Pillepich},
  {Rodriguez-Gomez}, {Torrey}, {Genel}, {Vogelsberger}, {Pakmor}, {Marinacci},
  {Weinberger}, {Kelley}, {Lovell}, {Diemer}, \&
  {Hernquist}}]{Nelson+Springel+2019}
{Nelson}, D., {Springel}, V., {Pillepich}, A., {et~al.} 2019, Computational
  Astrophysics and Cosmology, 6, 2, \dodoi{10.1186/s40668-019-0028-x}

\bibitem[{{{\"O}lmez} {et~al.}(2010){{\"O}lmez}, {Mandic}, \&
  {Siemens}}]{cosmic_string_spectrum}
{{\"O}lmez}, S., {Mandic}, V., \& {Siemens}, X. 2010, \prd, 81, 104028,
  \dodoi{10.1103/PhysRevD.81.104028}

\bibitem[{{Perera} {et~al.}(2019){Perera}, {DeCesar}, {Demorest}, {Kerr},
  {Lentati}, {Nice}, {Os{\l}owski}, {Ransom}, {Keith}, {Arzoumanian}, {Bailes},
  {Baker}, {Bassa}, {Bhat}, {Brazier}, {Burgay}, {Burke-Spolaor}, {Caballero},
  {Champion}, {Chatterjee}, {Chen}, {Cognard}, {Cordes}, {Crowter}, {Dai},
  {Desvignes}, {Dolch}, {Ferdman}, {Ferrara}, {Fonseca}, {Goldstein},
  {Graikou}, {Guillemot}, {Hazboun}, {Hobbs}, {Hu}, {Islo}, {Janssen},
  {Karuppusamy}, {Kramer}, {Lam}, {Lee}, {Liu}, {Luo}, {Lyne}, {Manchester},
  {McKee}, {McLaughlin}, {Mingarelli}, {Parthasarathy}, {Pennucci}, {Perrodin},
  {Possenti}, {Reardon}, {Russell}, {Sanidas}, {Sesana}, {Shaifullah},
  {Shannon}, {Siemens}, {Simon}, {Spiewak}, {Stairs}, {Stappers}, {Swiggum},
  {Taylor}, {Theureau}, {Tiburzi}, {Vallisneri}, {Vecchio}, {Wang}, {Zhang},
  {Zhang}, {Zhu}, \& {Zhu}}]{ipta_dr2}
{Perera}, B.~B.~P., {DeCesar}, M.~E., {Demorest}, P.~B., {et~al.} 2019, \mnras,
  490, 4666, \dodoi{10.1093/mnras/stz2857}

\bibitem[{{Phinney}(2001)}]{Phinney2001}
{Phinney}, E.~S. 2001, arXiv e-prints, astro.
\newblock \doarXiv{astro-ph/0108028}

\bibitem[{{Rajagopal} \& {Romani}(1995)}]{1995ApJ...446..543R}
{Rajagopal}, M., \& {Romani}, R.~W. 1995, \apj, 446, 543,
  \dodoi{10.1086/175813}

\bibitem[{{Rodriguez-Gomez} {et~al.}(2015){Rodriguez-Gomez}, {Genel},
  {Vogelsberger}, {Sijacki}, {Pillepich}, {Sales}, {Torrey}, {Snyder},
  {Nelson}, {Springel}, {Ma}, \& {Hernquist}}]{Rodriguez-Gomez+2015}
{Rodriguez-Gomez}, V., {Genel}, S., {Vogelsberger}, M., {et~al.} 2015, \mnras,
  449, 49, \dodoi{10.1093/mnras/stv264}

\bibitem[{{Romano} \& {Cornish}(2017)}]{2017LRR....20....2R}
{Romano}, J.~D., \& {Cornish}, N.~J. 2017, Living Reviews in Relativity, 20, 2,
  \dodoi{10.1007/s41114-017-0004-1}

\bibitem[{{Romano} {et~al.}(2020){Romano}, {Hazboun}, {Siemens}, \&
  {Archibald}}]{romano_acor_evn}
{Romano}, J.~D., {Hazboun}, J.~S., {Siemens}, X., \& {Archibald}, A.~M. 2020,
  arXiv e-prints, arXiv:2012.03804.
\newblock \doarXiv{2012.03804}

\bibitem[{{Rosado} {et~al.}(2015){Rosado}, {Sesana}, \&
  {Gair}}]{rosado_forecast}
{Rosado}, P.~A., {Sesana}, A., \& {Gair}, J. 2015, \mnras, 451, 2417,
  \dodoi{10.1093/mnras/stv1098}

\bibitem[{{Sampson} {et~al.}(2015){Sampson}, {Cornish}, \&
  {McWilliams}}]{2015PhRvD..91h4055S}
{Sampson}, L., {Cornish}, N.~J., \& {McWilliams}, S.~T. 2015, \prd, 91, 084055,
  \dodoi{10.1103/PhysRevD.91.084055}

\bibitem[{{Sazhin}(1978)}]{PTA_sazhin}
{Sazhin}, M.~V. 1978, Soviet Astronomy, 22, 36

\bibitem[{{Sesana} {et~al.}(2016){Sesana}, {Shankar}, {Bernardi}, \&
  {Sheth}}]{Sesana+201603}
{Sesana}, A., {Shankar}, F., {Bernardi}, M., \& {Sheth}, R.~K. 2016, \mnras,
  463, L6, \dodoi{10.1093/mnrasl/slw139}

\bibitem[{{Sesana} {et~al.}(2008){Sesana}, {Vecchio}, \&
  {Colacino}}]{2008MNRAS.390..192S}
{Sesana}, A., {Vecchio}, A., \& {Colacino}, C.~N. 2008, \mnras, 390, 192,
  \dodoi{10.1111/j.1365-2966.2008.13682.x}

\bibitem[{{Siemens} {et~al.}(2013){Siemens}, {Ellis}, {Jenet}, \&
  {Romano}}]{os_scaling_laws_siemens}
{Siemens}, X., {Ellis}, J., {Jenet}, F., \& {Romano}, J.~D. 2013, Classical and
  Quantum Gravity, 30, 224015, \dodoi{10.1088/0264-9381/30/22/224015}

\bibitem[{{Simon} \& {Burke-Spolaor}(2016)}]{2016ApJ...826...11S}
{Simon}, J., \& {Burke-Spolaor}, S. 2016, \apj, 826, 11,
  \dodoi{10.3847/0004-637X/826/1/11}

\bibitem[{{Taylor} {et~al.}(2018){Taylor}, {Baker}, {Hazboun}, {Simon}, \&
  {Vigeland}}]{enterprise_ext}
{Taylor}, S.~R., {Baker}, P.~T., {Hazboun}, J.~S., {Simon}, J.~J., \&
  {Vigeland}, S.~J. 2018, enterprise extensions.
\newblock \url{https://github.com/nanograv/enterprise_extensions}

\bibitem[{{Taylor} {et~al.}(2013){Taylor}, {Gair}, \&
  {Lentati}}]{2013PhRvD..87d4035T}
{Taylor}, S.~R., {Gair}, J.~R., \& {Lentati}, L. 2013, \prd, 87, 044035,
  \dodoi{10.1103/PhysRevD.87.044035}

\bibitem[{{Taylor} {et~al.}(2017){Taylor}, {Simon}, \&
  {Sampson}}]{2017PhRvL.118r1102T}
{Taylor}, S.~R., {Simon}, J., \& {Sampson}, L. 2017, prl, 118, 181102,
  \dodoi{10.1103/PhysRevLett.118.181102}

\bibitem[{{Taylor} {et~al.}(2016){Taylor}, {Vallisneri}, {Ellis}, {Mingarelli},
  {Lazio}, \& {van Haasteren}}]{2016ApJ...819L...6T}
{Taylor}, S.~R., {Vallisneri}, M., {Ellis}, J.~A., {et~al.} 2016, \apjl, 819,
  L6, \dodoi{10.3847/2041-8205/819/1/L6}

\bibitem[{{Taylor} {et~al.}(2020){Taylor}, {van Haasteren}, \&
  {Sesana}}]{2020PhRvD.102h4039T}
{Taylor}, S.~R., {van Haasteren}, R., \& {Sesana}, A. 2020, \prd, 102, 084039,
  \dodoi{10.1103/PhysRevD.102.084039}

\bibitem[{{Tiburzi} {et~al.}(2016){Tiburzi}, {Hobbs}, {Kerr}, {Coles}, {Dai},
  {Manchester}, {Possenti}, {Shannon}, \& {You}}]{tiburzi_spatial_corr}
{Tiburzi}, C., {Hobbs}, G., {Kerr}, M., {et~al.} 2016, \mnras, 455, 4339,
  \dodoi{10.1093/mnras/stv2143}

\bibitem[{{Vallisneri}(2020)}]{libstempo}
{Vallisneri}, M. 2020, {libstempo: Python wrapper for Tempo2}.
\newblock \doeprint{2002.017}

\bibitem[{{Vallisneri} {et~al.}(2020){Vallisneri}, {Taylor}, {Simon},
  {Folkner}, {Park}, {Cutler}, {Ellis}, {Lazio}, {Vigeland}, {Aggarwal},
  {Arzoumanian}, {Baker}, {Brazier}, {Brook}, {Burke-Spolaor}, {Chatterjee},
  {Cordes}, {Cornish}, {Crawford}, {Cromartie}, {Crowter}, {DeCesar},
  {Demorest}, {Dolch}, {Ferdman}, {Ferrara}, {Fonseca}, {Garver-Daniels},
  {Gentile}, {Good}, {Hazboun}, {Holgado}, {Huerta}, {Islo}, {Jennings},
  {Jones}, {Jones}, {Kaplan}, {Kelley}, {Key}, {Lam}, {Levin}, {Lorimer},
  {Luo}, {Lynch}, {Madison}, {McLaughlin}, {McWilliams}, {Mingarelli}, {Ng},
  {Nice}, {Pennucci}, {Pol}, {Ransom}, {Ray}, {Siemens}, {Spiewak}, {Stairs},
  {Stinebring}, {Stovall}, {Swiggum}, {van Haasteren}, {Witt}, \&
  {Zhu}}]{bayesephem}
{Vallisneri}, M., {Taylor}, S.~R., {Simon}, J., {et~al.} 2020, \apj, 893, 112,
  \dodoi{10.3847/1538-4357/ab7b67}

\bibitem[{Vaskonen \& Veerm\"ae(2021)}]{ng12p5_pbh_1}
Vaskonen, V., \& Veerm\"ae, H. 2021, Phys. Rev. Lett., 126, 051303,
  \dodoi{10.1103/PhysRevLett.126.051303}

\bibitem[{{Verbiest} {et~al.}(2016){Verbiest}, {Lentati}, {Hobbs}, {van
  Haasteren}, {Demorest}, {Janssen}, {Wang}, {Desvignes}, {Caballero}, {Keith},
  {Champion}, {Arzoumanian}, {Babak}, {Bassa}, {Bhat}, {Brazier}, {Brem},
  {Burgay}, {Burke-Spolaor}, {Chamberlin}, {Chatterjee}, {Christy}, {Cognard},
  {Cordes}, {Dai}, {Dolch}, {Ellis}, {Ferdman}, {Fonseca}, {Gair},
  {Garver-Daniels}, {Gentile}, {Gonzalez}, {Graikou}, {Guillemot}, {Hessels},
  {Jones}, {Karuppusamy}, {Kerr}, {Kramer}, {Lam}, {Lasky}, {Lassus},
  {Lazarus}, {Lazio}, {Lee}, {Levin}, {Liu}, {Lynch}, {Lyne}, {Mckee},
  {McLaughlin}, {McWilliams}, {Madison}, {Manchester}, {Mingarelli}, {Nice},
  {Os{\l}owski}, {Palliyaguru}, {Pennucci}, {Perera}, {Perrodin}, {Possenti},
  {Petiteau}, {Ransom}, {Reardon}, {Rosado}, {Sanidas}, {Sesana}, {Shaifullah},
  {Shannon}, {Siemens}, {Simon}, {Smits}, {Spiewak}, {Stairs}, {Stappers},
  {Stinebring}, {Stovall}, {Swiggum}, {Taylor}, {Theureau}, {Tiburzi},
  {Toomey}, {Vallisneri}, {van Straten}, {Vecchio}, {Wang}, {Wen}, {You},
  {Zhu}, \& {Zhu}}]{ipta_dr1}
{Verbiest}, J.~P.~W., {Lentati}, L., {Hobbs}, G., {et~al.} 2016, \mnras, 458,
  1267, \dodoi{10.1093/mnras/stw347}

\bibitem[{{Vigeland} {et~al.}(2018){Vigeland}, {Islo}, {Taylor}, \&
  {Ellis}}]{NM_OS}
{Vigeland}, S.~J., {Islo}, K., {Taylor}, S.~R., \& {Ellis}, J.~A. 2018, \prd,
  98, 044003, \dodoi{10.1103/PhysRevD.98.044003}

\bibitem[{{Vigeland} \& {Siemens}(2016)}]{sarah_forecast}
{Vigeland}, S.~J., \& {Siemens}, X. 2016, \prd, 94, 123003,
  \dodoi{10.1103/PhysRevD.94.123003}

\bibitem[{{Vogelsberger} {et~al.}(2014{\natexlab{a}}){Vogelsberger}, {Genel},
  {Springel}, {Torrey}, {Sijacki}, {Xu}, {Snyder}, {Bird}, {Nelson}, \&
  {Hernquist}}]{Vogelsberger+2014a}
{Vogelsberger}, M., {Genel}, S., {Springel}, V., {et~al.} 2014{\natexlab{a}},
  nat, 509, 177, \dodoi{10.1038/nature13316}

\bibitem[{{Vogelsberger} {et~al.}(2014{\natexlab{b}}){Vogelsberger}, {Genel},
  {Springel}, {Torrey}, {Sijacki}, {Xu}, {Snyder}, {Nelson}, \&
  {Hernquist}}]{Vogelsberger+2014b}
---. 2014{\natexlab{b}}, \mnras, 444, 1518, \dodoi{10.1093/mnras/stu1536}

\end{thebibliography}

\end{document}